\batchmode
\makeatletter
\def\input@path{{\string"C:/_cvut/habilitace/arxiv_org/PCT paper/\string"}}
\makeatother
\documentclass{IEEEtran}
\usepackage[latin9]{inputenc}
\usepackage{color}
\usepackage{array}
\usepackage{rotfloat}
\usepackage{algorithm2e}
\usepackage{amsmath}
\usepackage{amssymb}
\usepackage{graphicx}

\makeatletter

\providecommand{\tabularnewline}{\\}

\ifCLASSINFOpdf
 
\else

\fi

\usepackage{url}
\usepackage{breakurl}
\usepackage[breaklinks]{hyperref}
\RestyleAlgo{ruled}
\LinesNumbered

\makeatother

\begin{document}

\title{Employment of Multiple Algorithms for Optimal Path-based Test Selection
Strategy}

\author{Miroslav Bures and Bestoun S. Ahmed \thanks{M. Bures, Software Testing Intelligent Lab (STILL), Department of
Computer Science, Faculty of Electrical Engineering Czech Technical
University, Karlovo nam. 13, 121 35 Praha 2, Czech Republic, (email:
buresm3@fel.cvut.cz)} \thanks{B. Ahmed, Software Testing Intelligent Lab (STILL), Department of
Computer Science, Faculty of Electrical Engineering Czech Technical
University, Karlovo nam. 13, 121 35 Praha 2, Czech Republic, (email:
albeybes@fel.cvut.cz)} }
\maketitle
\begin{abstract}
Executing various sequences of system functions in a system under
test represents one of the primary techniques in software testing.
The natural way to create effective, consistent and efficient test
sequences is to model the system under test and employ an algorithm
to generate the tests that satisfy a defined test coverage criterion.
Several criteria of test set optimality can be defined. In addition,
to optimize the test set from an economic viewpoint, the priorities
of the various parts of the system model under test must be defined.
Using this prioritization, the test cases exercise the high priority
parts of the system under test more intensely than those with low
priority. Evidence from the literature and our observations confirm
that finding a universal algorithm that produces an optimal test set
for all test coverage and test set optimality criteria is a challenging
task. Moreover, for different individual problem instances, different
algorithms provide optimal results. In this paper, we present a path-based
strategy to perform optimal test selection. The strategy first employs
a set of current algorithms to generate test sets; then, it assesses
the optimality of each test set by the selected criteria, and finally,
chooses the optimal test set. The experimental results confirm the
validity and usefulness of this strategy. For individual instances
of 50 system under test models, different algorithms provided optimal
results; these results varied by the required test coverage level,
the size of the priority parts of the model, and the selected test
set optimality criteria.
\end{abstract}


\section{Introduction}

\IEEEPARstart{T}{he} natural way to construct a test case is to chain
a sequence of specific calls to various functions of the system under
test (SUT). Whether designing a method flow, or API calls are used
for an integration test, or a test scenario is designed for a manual
business end-to-end test, following a systematic approach that generates
consistent and effective test sequences is essential. The field of
model-based testing provides a solution for this issue through which
we first model a particular SUT process or workflow in a suitable
notation and then use an appropriate algorithm to generate the flows
(i.e., path-based test cases).

To generate the path-based test cases systematically and consistently,
a SUT model based on a directed graph is used \cite{offutt2008introduction}.
Several algorithms have been presented (e.g., \cite{dwarakanath2014minimum,li2012better,arora2017synthesizing,sayyari2015automated,hoseini2014automatic,shirole2013uml})
to solve this problem. However, based on both evidence from the literature
and our experiments while developing new algorithms to solve this
problem, it is challenging task to find a universal algorithm that
can generate an optimal test set for all instances. Not only do individual
problem instances (particular SUT models) differ but also different
test set optimality criteria can be formulated \cite{li2012better,offutt2008introduction,li2009experimental}.
A significant finding here addresses the possibility for creating
a universal algorithm that satisfies multiple optimality criteria.
This task is complicated and must consider the different test coverage
criteria that have been defined, which span a range from All Node
Coverage to All Path Coverage, and individual algorithms differ in
their ability to produce test sets that satisfy these different criteria
\cite{shirole2013uml}.

The complexity of the problem increases when individual parts of the
SUT model should be tested at different priority levels. For instance,
consider a complex workflow in an information system that must be
covered by path-based test scenarios. Only selected parts of the workflow
require coverage by high-intensity test scenarios, while for the remaining
parts, lightweight tests are sufficient to optimize the test set and
reduce the testing costs. The priorities can captured by the test
requirements \cite{offutt2008introduction,li2012better} or by defining
edge weights in the model \cite{bures2017prioritized}. However, to
reflect the priorities captured by edge weights when generating test
cases, alternative strategies must be defined because the current
algorithms provide near optimum results only for non-prioritized SUT
models and can be suboptimal when solving this type of problem for
prioritized SUT models.

To address the issues described above, in this study, we employ an
approach based on combining current algorithms, including both our
own work in this area \cite{bures2017prioritized} and selected algorithms
previously published in the literature \cite{li2012better,offutt2008introduction}.
The strategy, which includes these algorithms, relies on input from
the tester as follows. The tester first creates a SUT model, defines
the priority parts of the model, and specifies the test coverage criteria.
Then, the tester selects the test set optimality criteria from a set
of options (details are provided in Section \ref{subsec:Selection-of-the-best-test-set}).
This strategy uses all the algorithms to generate different test sets
based on the SUT model. Then, based on the test set optimality criteria,
the best test set is selected and provided to the test analyst. We
implemented this test case generation strategy in the latest version
of the experimental Oxygen Model-based Testing platform\footnote{http://still.felk.cvut.cz/oxygen/}
developed by the STILL group. In this paper, we present the details
of this strategy and its results for 50 SUT models using Edge Coverage
and Edge-Pair Coverage criteria and for 16 different test set optimality
criteria (including an optimality function and a sequence-selection
strategy composed from additional test set optimality indicators).
These data can also be used to compare the test sets produced by the
algorithms.

The paper is organized as follows. Section \ref{sec:Problem-Definition}
defines the problem; then, it provides an overview of the test coverage
criteria used to determine the intensity of the test set, and finally,
it discusses possible test set optimality criteria. Section \ref{sec:Selection-of-The}
provides the details of the process for selecting an optimal test
set based on the optimality criteria. Section \ref{sec:Experiments}
presents the experimental method and the acquired data. Section \ref{sec:Discussion}
discusses the results and Section \ref{sec:Threats-to-Validity} analyzes
possible threats to validity. Section \ref{sec:Related-Work} summarizes
the relevant related work. Finally, Section \ref{sec:Conclusion}
concludes this paper.

\section{\label{sec:Problem-Definition}Problem Definition}

As mentioned previously, the strategy presented in this paper takes
a SUT model as input. Here, the SUT process is modeled as a directed
graph $G=(N,E)$, where $N$ is a set of nodes, $N\neq\emptyset$,
and $E$ is a set of edges. $E$ is a subset of $N\times N$. In the
model we define one start node $n_{s}\in N$. The set $N_{e}\subseteq N$
contains the end nodes of the graph, and $N_{e}\neq\emptyset$ \cite{offutt2008introduction}. 

The SUT functions and decision points are mapped to $G$ depending
on the level of abstraction. In addition, the SUT layer for which
we prepare test cases plays an essential role in the modeling. As
an example, we can provide data-flow testing at the code level or
design an end-to-end (E2E) high-level business-process test set. More
information about this topic appears in \ref{sec:Discussion}.

The test case $t$ is a sequence of nodes $n_{1},n_{2},..,n_{n}$,
with a sequence of edges $e_{1},e{}_{2},..,e_{n-1}$, where $e_{i}=(n_{i,}n_{i+1})$,
$e_{i}\in E$, $n_{i},\in N$, $n_{i\text{+1}}\in N$. The test case
$t$ starts with the start node $n_{s}$ ($n_{1}=n_{s}$) and ends
with a $G$ end node ($n_{n}\in N_{e}$) . We can denote the test
case $t$ as a either sequence of nodes $n_{1},n_{2},..,n_{n}$, or
a sequence of edges $e_{1},e{}_{2},..,e_{n-1}$. The test set $T$
is a set of test cases. 

To determine the required the test coverage, we define a set of test
requirements $R$. Generally, a test requirement is a path in $G$
that must be a sub-path of at least one test case $t\in T$. The test
requirements can be used either to (1) define the general intensity
of the test cases or (2) to express which parts of the SUT model $G$
are considered as priorities to be covered by test cases. 

The fact that the test requirements can be used either to determine
the overall intensity of the test set $T$ or to express which parts
of the SUT model $G$ should be tested at a higher priority leads
us to adopt an alternative definition of the SUT model. This definition
supports formulation of algoriths which allow determining testing
intensity and expressing priorities in parallel \cite{bures2017prioritized}.
Moreover, it uses a multigraph instead of a graph as a SUT model,
which gives test analysts more flexibility when modeling SUT processes
(this issue is discussed further in Section \ref{sec:Discussion}).
In addition, using more priority levels is natural in the software
development process \cite{achimugu2014systematic}; using test requirements
for prioritization results in algorithms being able to work with two
priority levels only, which could restrict the development of further
and possibly more effective algorithms. 

Our alternative definition of the SUT model is as follows. We model
a SUT process as a weighted multigraph $\mathfrak{\mathcal{G}}=(N,E,s,t)$,
where $N$ is a set of nodes, $N\neq\emptyset$, and $E$ is a set
of edges. Here, $s:\,E\rightarrow N$ assigns each edge to its source
node and $t:\,E\rightarrow N$ assigns each edge to its target node.
We define one start node $n_{s}\in N$. The set $N_{e}\subseteq N$
contains the end nodes of the multigraph, $N_{e}\neq\emptyset$. For
each edge $e\in E$ (resp. node $n\in N$), a $priority(e)$ (resp.
$priority(n)$) is defined, where $priority(e)\in\{high,medium,low\}$
and $priority(n)\in\{high,medium,low\}$. When $p$ is not defined,
the default $low$ value is used. $E_{h}$ is a set of high-priority
edges; $E_{m}$ is a set of medium-priority edges; and $E_{l}$ is
a set of low-priority edges, where $E_{h}\cup E_{m}\cup E_{l}=E$,
$E_{h}\cap E_{m}=\emptyset$, $E_{m}\cap E_{l}=\emptyset$, $E_{h}\cap E_{l}=\emptyset$.

Priority $p$ reflects the importance of the edge to be tested. The
test analyst determines the priority based on a risk prioritization
technique \cite{van2013tmap} or a technique that combines risk assessment
with information regarding the internal complexity of the SUT or the
presence of defects in previous SUT versions \cite{achimugu2014systematic}.
To determine the intensity of the test set $T$, test coverage criteria
are used.

\subsection{\label{subsec:Test-coverage-criteria}Test Coverage Criteria}

Several different test coverage criteria have been defined for $T$.
For instance, \textit{All Edge Coverage} (or\textit{ Edge Coverage})
requires each edge $e\in E$ to be present in the test set $T$ minimally
once. Alternatively, \textit{All Node Coverage} requires each node
$n\in N$ to be present in test set $T$ at least once. To satisfy
the \textit{Edge-Pair Coverage} criterion, the test set $T$ must
contain each possible pair of adjacent edges in $G$ \cite{offutt2008introduction}. 

The\textit{ All Paths Coverage }(or Complete Path Coverage) requires
that all possible paths in $G$, starting from $n_{s}$ and ending
at any node of $N_{e}$, be present in test set $T$. Such a test
set can contain considerable redundancy. To reduce this redundancy,
the \textit{Prime Path Coverage} criterion is used. To satisfy the
\textit{Prime Path Coverage} criterion, each reachable prime path
in $G$ must be a sub-path of a test case $t\in T$. A path $p$ from
$e_{1}$ to $e_{2}$is prime if (1) $p$ is simple, and (2) $p$ is
not a sub-path of any other simple path in $G$. A path $p$ is simple
when no node $n\in N$ is present more than once in $p$ (i.e., $p$
does not contain any loops); the only exception is $e_{1}$ and $e_{2}$,
which can be identical (in other words, $p$ itself can be a loop)
\cite{offutt2008introduction}.

Sorted by the intensity of the test cases, \textit{\textcolor{black}{All
Node Coverage }}is the weakest option, followed by \textit{All Edge
Coverage}, \textit{Edge-Pair Coverag}\textit{\textcolor{black}{e}}\textit{
}\textcolor{black}{and }\textit{\textcolor{black}{Prime Path Coverage}}\textcolor{black}{.
The}\textit{\textcolor{black}{{} All Paths Coverage}}\textit{\textcolor{red}{{}
}}\textcolor{black}{lies at the other end of the spectrum \cite{offutt2008introduction},
}because it implies the most intense test cases. However, due to the
high number of test case steps, this option is not practicable in
most software development projects. This problem can be also faced
by \textit{Prime Path Coverage}: for many routine process-testing
tasks, this level of test coverage can be too extensive.

Test coverage criteria can also be specified by the \textit{Test Depth
Level (TDL)} \cite{koomen2013tmap}. $TDL=1$ when $\forall e\in E$
the edge $e$ appears at least once in at least one test case $t\in T$.
$TDL=x$ when the test set $T$ satisfies the following conditions:
For each node $n\in N$, $P_{n}$ is a set of all possible paths in
$G$ starting with an edge incoming to the decision point $n$, followed
by a sequence of $x-1$ edges outgoing from the node $n$. Then, $\forall n\in N$
the test cases cases in test set $T$ contain all paths from $P_{n}$.
When $TDL=1$, it is equivalent to \textit{All Edge Coverage}, and
when $TDL=2$, it is equivalent to \textit{Edge-Pair Coverage}. All
the coverage criteria in this section are defined in the same way
for $G$ as well as for $\mathfrak{\mathcal{G}}$.

To determine the testing priority in selected parts of the SUT processes,
we define a \textit{Priority Level (PL)} for $\mathfrak{\mathcal{G}}$.
$PL\in\{high,medium\}$. $PL=high$ when $\forall e\in E_{h}$ the
edge $e$ is present at least once in at least one test case $t\in T$.
Further, $PL=medium$ when $\forall e\in E_{h}\cup E_{m}$ the edge
$e$ is present at least once in at least one test case $t\in T$.
When a test set $T$ satisfies the \textit{All Edge Coverage}, it
also satisfies \textit{PL}. 

In the test case generation strategies we evolve, the \textit{PL}
can be combined with yet another test coverage criteria such as \textit{TDL}
\cite{bures2017prioritized} or \textit{Prime Path Coverage} (refer
to Section \ref{subsec:RSC-Algorithm}). In these cases, \textit{PL}
reduces the test coverage by \textit{TDL }or \textit{Prime Paths Coverage}
to only the $\mathfrak{\mathcal{G}}$ parts, which are defined as
the priority. It allows optimizing the test cases to exercise only
the priority parts of the SUT processes or workflows.

\subsection{\label{sec:Test-Set-Optimality-Criteria}Test Set Optimality Criteria}

Various optimality criteria for $T$ have been discussed in the literature
(e.g., \cite{li2012better,li2009experimental}). Table \ref{tab:Test-set-optimality-criteria}
lists the optimality criteria used in this paper as defined for SUT
model $\mathfrak{\mathcal{G}}$. Parts of these criteria can also
be defined for $G$ which is captured in Table \ref{tab:Test-set-optimality-criteria}
in the column \textquotedblleft Applicable to.\textquotedblright{}

\begin{table*}
\scriptsize

\caption{\label{tab:Test-set-optimality-criteria}Test set $T$ optimality
criteria}

\centering{}%
\begin{tabular}{|c|>{\centering}p{8.5cm}|c|}
\hline 
Optimality criterion &
Description &
Applicable to\tabularnewline
\hline 
\hline 
$\mid T\mid$  &
Number of test cases in the test set &
$\mathfrak{\mathcal{G}}$ and $G$\tabularnewline
\hline 
$edges(T)$ &
Total number of edges in the test cases of a test set $T$, edges
can repeat &
$\mathfrak{\mathcal{G}}$ and $G$\tabularnewline
\hline 
$edges_{h}(T)$ &
Total number of edges of priority $high$ in the test cases of a test
set $T$, edges can repeat &
$\mathfrak{\mathcal{G}}$\tabularnewline
\hline 
$edges_{m}(T)$ &
Total number of edges of priority $high$ and $medium$ in the test
cases of a test set $T$, edges can repeat &
$\mathfrak{\mathcal{G}}$\tabularnewline
\hline 
$uedges(T)$ &
Total number of unique edges in the test cases of a test set $T$ &
$\mathfrak{\mathcal{G}}$ and $G$\tabularnewline
\hline 
$uedges_{h}(T)$ &
Total number of unique edges of priority $high$ in the test cases
of a test set $T$ &
$\mathfrak{\mathcal{G}}$\tabularnewline
\hline 
$uedges_{m}(T)$ &
Total number of unique edges of priority $high$ and $medium$ in
the test cases of a test set $T$ &
$\mathfrak{\mathcal{G}}$\tabularnewline
\hline 
$nodes(T)$ &
Total number of nodes in the test cases of a test set $T$, nodes
can repeat &
$\mathfrak{\mathcal{G}}$ and $G$\tabularnewline
\hline 
$unodes(T)$ &
Total number of unique nodes in the test cases of a test set $T$ &
$\mathfrak{\mathcal{G}}$\tabularnewline
\hline 
$er(T)=\frac{uedges(T)}{\mid E\mid}.100\%$ &
Ratio of unique edges contained in the test cases of a test set $T$.
A lower value of $er(T)$ means more optimal test set, because less
unique edges are present in the test cases and these unique edges
can represent extra costs for preparation of the detailed test scenarios. &
$\mathfrak{\mathcal{G}}$ and $G$\tabularnewline
\hline 
$e_{h}(T)=\frac{edges_{h}(T)}{edges(T)}.100\%$ &
Ratio of edges of priority $high$ and all edges in the test cases
of a test set $T$. A higher value of $e_{h}(T)$ means more optimal
test set, because less edges which do not have priority $high$ (thus
are not necessary to test) are present in the test cases. &
$\mathfrak{\mathcal{G}}$\tabularnewline
\hline 
$e_{m}(T)=\frac{edges_{m}(T)}{edges(T)}.100\%$ &
Ratio of edges of priority $high$ and $medium$ and all edges in
the test cases of a test set $T$. A higher value of $e_{h}(T)$ means
more optimal test set, because less edges which do not have priority
$high$ and $medium$ (thus are not necessary to test) are present
in the test cases. &
$\mathfrak{\mathcal{G}}$\tabularnewline
\hline 
$ue_{h}(T)=\frac{uedges_{h}(T)}{edges(T)}.100\%$ &
The same as $e_{h}(T)$, only unique edges are taken in acount &
$\mathfrak{\mathcal{G}}$\tabularnewline
\hline 
$ue_{m}(T)=\frac{uedges_{m}(T)}{edges(T)}.100\%$ &
The same as $e_{m}(T)$, only unique edges are taken in acount &
$\mathfrak{\mathcal{G}}$\tabularnewline
\hline 
\end{tabular}
\end{table*}

Individual test set optimality criteria can be combined. In this paper,
we explore two possible methods: combining the optimality criteria
to a formula and evaluating the test set $T$ using a sequence of
criteria. The following section provides more detail on the selection
of the optimal test set and how these methods can be used successfully. 

\section{\label{sec:Selection-of-The}Selecting an Optimal Test Set}

To obtain an optimal test set $T$ for SUT model $\mathfrak{\mathcal{G}}$
along with test coverage and test set optimality criteria, we conduct
a sequence of three main steps. First, we select a set of algorithms
and their suitable input parameters to generate the $T$ for $\mathfrak{\mathcal{G}}$,
the test coverage, and the test set optimality criteria. Then, we
run these selected algorithms to produce the test sets $T_{1}..T_{m}$.
Finally, we analyze the test sets $T_{1}..T_{m}$ and select the test
set $T$ that has the best value of the optimality criteria. The \textbf{inputs}
to this process are as follows:
\begin{enumerate}
\item SUT model $\mathfrak{\mathcal{G}}$
\item Test coverage criteria from the following options:
\begin{enumerate}
\item Test intensity from the following options: \textit{Edge Coverage},
\textit{Edge-Pair Coverage}, $TDL$ (where $TDL>2$, because $TDL=1$
is equivalent to Edge Coverage and $TDL=2$ is equivalent to Edge-Pair
coverage), and \textit{Prime Path Coverage.}
\item Coverage of the $\mathfrak{\mathcal{G}}$ priority parts by \textit{Priority
Level (PL) }as defined in Section \ref{subsec:Test-coverage-criteria}.
\end{enumerate}
\item Test set optimality criterion from the options defined in Section
\ref{sec:Test-Set-Optimality-Criteria} and Table \ref{tab:Test-set-optimality-criteria}.
\end{enumerate}
The \textbf{output} of the process is an optimal test set $T$, that
satisfies the test coverage criterion.

\subsection{Included Algorithms}

In the described strategy, we use the algorithms listed in Table \ref{tab:Algorithms-used-for}.

\begin{table*}
\scriptsize

\caption{\label{tab:Algorithms-used-for}Algorithms used to generate $T$}

\centering{}%
\begin{tabular}{|c|>{\centering}p{4cm}|>{\centering}p{5cm}|>{\centering}p{3cm}|}
\hline 
Code &
Name &
SUT model &
Reference\tabularnewline
\hline 
\hline 
PCT &
Process Cycle Test &
$G$ &
Koomen et al.\cite{koomen2013tmap}, Bures \cite{bures2015pctgen}\tabularnewline
\hline 
PPT &
Prioritized Process Test &
$\mathfrak{\mathcal{G}}$ &
Bures et al. \cite{bures2017prioritized}\tabularnewline
\hline 
BF &
Brute Force Solution &
$G$, set of test requirements $R$ &
Li, Li and Offut \cite{li2012better}\tabularnewline
\hline 
SC &
Set-Covering Based Solution &
$G$, set of test requirements $R$ &
Li, Li and Offut \cite{li2012better}\tabularnewline
\hline 
PG &
Matching-Based Prefix Graph Solution &
$G$, set of test requirements $R$ &
Li, Li and Offut\textcolor{red}{{} }\cite{li2012better}\tabularnewline
\hline 
RSC &
Set-Covering Based Solution with Test Set Reduction &
$\mathfrak{\mathcal{G}}$ for the whole algorithm,$G$ and $R$ for
its SC part &
Specified in Section\textit{ }\textit{\textcolor{red}{\ref{subsec:RSC-Algorithm}}}\tabularnewline
\hline 
\end{tabular}
\end{table*}

We used the Oxygen Model-based Testing experimental platform\footnote{http://still.felk.cvut.cz/oxygen/}
(formerly PCTgen) \cite{bures2015pctgen} to implement the proposed
strategy. Our research team implemented process Cycle Test (PCT) and
Prioritized Process Test (PPT). We also implemented the Brute Force
Solution (BF) algorithm based on the pseudocode published by Li et
al. \cite{li2012better}. The implementations of the Set-Covering
Based Solution (SC) and Matching-Based Prefix Graph Solution (PG)
algorithms is based on the source code by Ammann and Offutt \cite{offuttoolsapplication}.
The Set-Covering Based Solution with Test Case Reduction (RSC) consists
of the Set-Covering Based Solution part and our implementation of
the test set reduction part (further specified in Section \ref{subsec:RSC-Algorithm}).

The role of the PCT algorithm is only to provide information on how
many test cases and test steps are in a test set $T$ for a particular
SUT model $\mathfrak{\mathcal{G}}$ when the SUT model parts are not
prioritized. 

\subsection{Test Set Generation Process}

The strategy to determine the optimal test set $T$ by the selected
test set optimality criterion consists of five main steps, which are
summarized in Algorithm \ref{alg:The-main-process-Algo}.

\begin{algorithm*}

\caption{\label{alg:The-main-process-Algo}The main process of test set generation
\textbf{ }}

\KwIn{$\mathfrak{\mathcal{G}}$, test coverage criteria, $PL\in\{high,medium\}$, test set optimality criterion}  
\KwOut{test set $T$}

Convert $\mathfrak{\mathcal{G}}$ to $G$ and $R$ for the BF, SC,
PG and RSC algorithms (refer to Section \ref{subsec:Conversion-of-})

Determine the set of algorithms that are suitable for generating the
T for $\mathfrak{\mathcal{G}}$ (refer to Section \ref{subsec:Conversion-of-})

Execute the selected set of algorithms with $\mathfrak{\mathcal{G}}$
(or with $G$ and $R$, which correspond to $\mathfrak{\mathcal{G}})$.
The output of this step are the test sets $T_{1}\ldots T_{m}$

Compute the values of single test set optimality criteria for $T_{1}\ldots T_{m}$
(refer to Table \ref{tab:Test-set-optimality-criteria} and Section
\ref{subsec:Selection-of-the-best-test-set}), which will be employed
in step 5.

Select the optimal $T$ of $T_{1}\ldots T_{m}$ as determined by the
test set optimality criterion (refer to Section \ref{subsec:Selection-of-the-best-test-set})
\end{algorithm*}

\begin{figure}
\begin{centering}
\includegraphics[width=9cm]{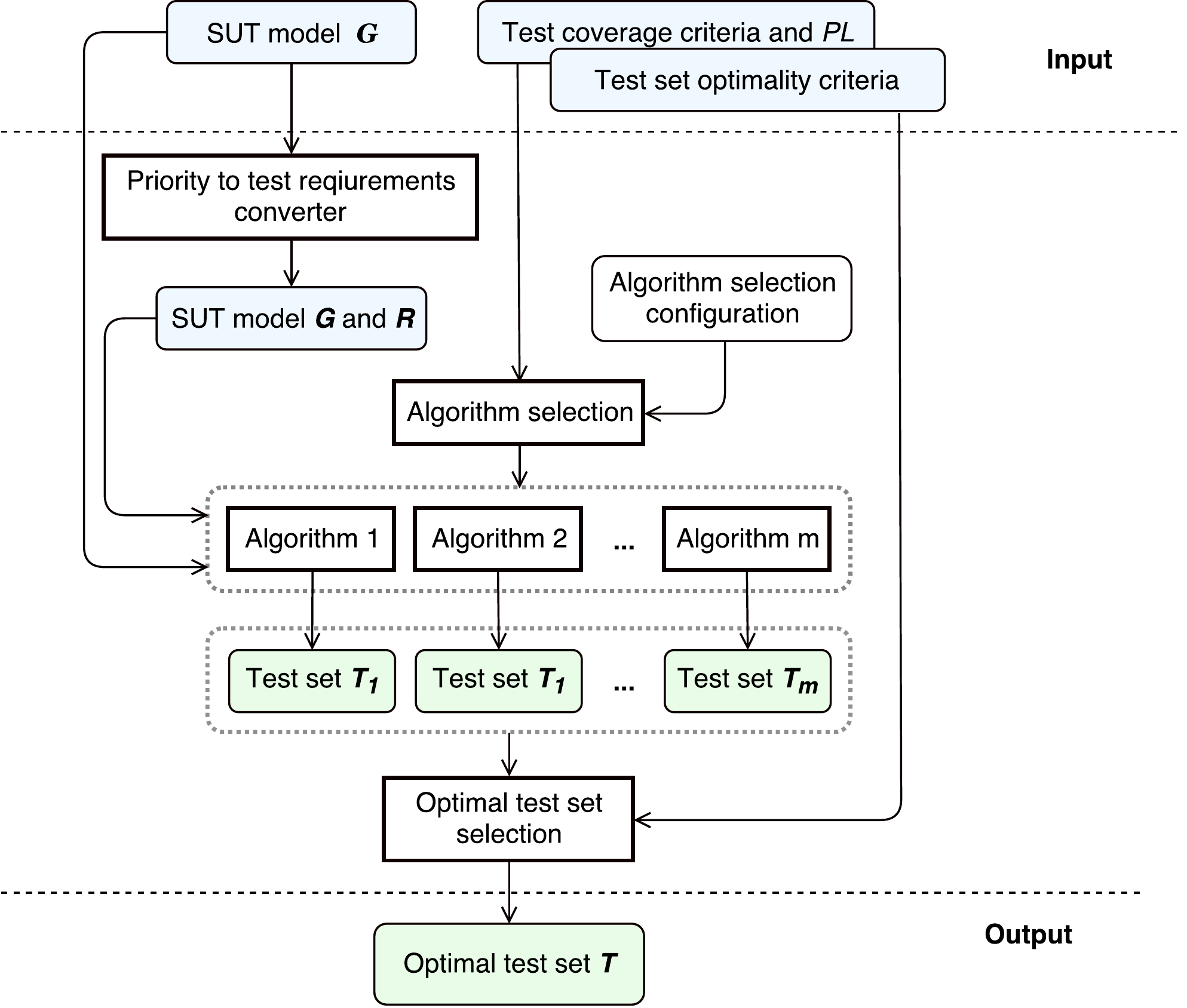}
\par\end{centering}
\caption{\label{fig:Overal-process-of}Main steps of the proposed test case
generation strategy }
\end{figure}

Figure \ref{fig:Overal-process-of} depicts the overall process. The
process inputs are marked in blue while the process outputs are marked
in green.

In the Oxygen platform, the $T$ is presented to the user, as well
as $T_{1}..T_{m}$. For each of $T_{1}..T_{m}$ , Oxygen also provides
the values of the optimality criteria. In the following subsections,
we explain these individual steps in more detail.

\subsubsection{\label{subsec:Conversion-of-}Conversion of $\mathfrak{\mathcal{G}}$
to $G$ and $R$}

For the BF, SC, PG and RSC algorithms, we need to convert graph $\mathfrak{\mathcal{G}}$
to graph $G$ and a set of test requirements, $R$. The multigraph$\mathfrak{\mathcal{G}}$
is equivalent to graph $G$, when (1) edge priorities ($priority(e)$)
and node priorities ($priority(n)$) \textcolor{black}{are not considered,
and (2) there are no parallel edges in }$\mathfrak{\mathcal{G}}$.
This conversion implies that when creating $\mathfrak{\mathcal{G}},$
no parallel edges can be used, which can restrict the modeling possibilities
that using a multigraph as a SUT process abstraction makes possible.
However, this restriction can be solved without losing the applicability
of the proposed strategy by modeling the parallel edges as graph nodes.

A set of test requirements $R$ is created by a method specified in
Table \ref{tab:Creation_of_Test_requirements}.

\begin{table*}
\scriptsize
\begin{centering}
\caption{\label{tab:Creation_of_Test_requirements}Method of creation of the
test requirements $R$ from $\mathfrak{\mathcal{G}}$}
\par\end{centering}
\centering{}%
\begin{tabular}{|>{\centering}p{3cm}|>{\centering}p{1.8cm}|>{\centering}p{3cm}|>{\centering}p{3cm}|}
\hline 
Test coverage: \textbf{test intensity} &
Method of $R$ creation &
$PL=high$ &
$PL=medium$\tabularnewline
\hline 
\hline 
\textit{Edge Coverage}

($TDL=1$) &
Atomic conversion &
$R$ is a set of all $G$ adjacent node pairs $e=(n_{i},n_{i+1})$
for each $e\in E_{h}$ &
$R$ is a set of all $G$ adjacent node pairs $e=(n_{i},n_{i+1})$
for each $e\in E_{h}\cup E_{m}$ \tabularnewline
\hline 
\textit{Edge Coverage}

($TDL=1$) &
Sequence conversion &
$R$ is a set of paths in $\mathfrak{\mathcal{G}}$, $priority(e)=high$for
each $e\in p\in R$ &
$R$ is a set of paths in $\mathfrak{\mathcal{G}}$, $priority(e)$$\in\{high,medium\}$for
each $e\in p\in R$\tabularnewline
\hline 
\textit{Edge-Pair Coverage} ($TDL=2$)  &
\multicolumn{3}{>{\centering}p{8cm}|}{A set $E_{pair}$ contains all possible pairs of adjacent edges of
$\mathcal{G}$. Then, $R$ is a set of all paths $(n_{i},n{}_{i+1},n_{i+2})$,
such that $e_{i}=(n_{i,},n_{i+1})$, $e_{i+1}=(n_{i+1,}n_{i+2})$
for each $(e_{i},e{}_{i+1})\in E_{pair}$. This process is not influenced
by $PL$.}\tabularnewline
\hline 
$TDL=x$,

$x>2$ &
\multicolumn{3}{>{\centering}p{8cm}|}{A set $E_{x}$ contains all possible paths of $\mathcal{G}$ consisting
of $x$ adjacent edges. Then, $R$ is a set of all paths $(n_{i},n{}_{i+1},n_{i+2},..,n_{x+1})$,
such that $e_{i}=(n_{i,},n_{i+1})$, $e_{i+1}=(n_{i+1,}n_{i+2})$,
... , $e_{x}=(n_{x,}n_{x+1})$ for each $(e_{i},e{}_{i+1},..,e_{x})\in E_{pair}$.
This process is not influenced by $PL$.}\tabularnewline
\hline 
\textit{Prime Path Coverage} &
\multicolumn{3}{>{\centering}p{8cm}|}{$R$ is a set of all possible prime paths in $\mathfrak{\mathcal{G}}$
(applies to BF, SC and PG). This process is not influenced by $PL$.}\tabularnewline
\hline 
\end{tabular}
\end{table*}

\subsubsection{\label{subsec:Selection-of-algorithms}Algorithms Selection}

Table \ref{tab:Selection-of-algorithms} specifies the process of
selecting algorithms by the specified test coverage criteria (algorithm
selection configuration as depicted in Fig. \ref{fig:Overal-process-of}).

\begin{table*}
\scriptsize
\begin{centering}
\caption{\label{tab:Selection-of-algorithms}Algorithm selection configuration}
\par\end{centering}
\centering{}%
\begin{tabular}{|>{\centering}p{3cm}|>{\centering}p{1.8cm}|>{\centering}p{3cm}|>{\centering}p{3cm}|}
\hline 
 &
\multicolumn{3}{c|}{Test set reduction: \textbf{coverage of priority parts of $\mathfrak{\mathcal{G}}$ }}\tabularnewline
\hline 
Test coverage: \textbf{test intensity} &
$PL=high$ &
$PL=medium$ &
not reduced by $PL$\tabularnewline
\hline 
\hline 
\textit{Edge Coverage}

($TDL=1$) &
PPT

RSC

BF

SC

PG &
PPT

RSC

BF

SC

PG &
PCT\tabularnewline
\hline 
\textit{Edge-Pair Coverage} ($TDL=2$)  &
PPT

RSC &
PPT

RSC &
PCT

BF

SC

PG\tabularnewline
\hline 
$TDL>2$ &
PPT &
PPT &
PCT

BF

SC

PG\tabularnewline
\hline 
\textit{Prime Path Coverage} &
RSC &
RSC &
BF

SC

PG\tabularnewline
\hline 
\end{tabular}
\end{table*}

For the \textit{Edge Coverage} case ($TDL=1$), the BF, SC and PG
algorithms reflect the edge priorities in \textbf{$\mathfrak{\mathcal{G}}$
}via a set of test requirements, $R$, generated from $G$ (refer
to Table \ref{tab:Creation_of_Test_requirements}). In both conversion
types, the Atomic and the Sequence conversions are used for each of
these algorithms. In contrast, PPT and RSC work directly with the
edge priorities in \textbf{$\mathfrak{\mathcal{G}}$ }(\cite{bures2017prioritized}\textcolor{blue}{{}
}and Section \ref{subsec:RSC-Algorithm}). 

For the \textit{Edge-Pair Coverage} case ($TDL=2$), the PPT and RSC
algorithms are comparable candidates when the $PL$ criterion reduces
the test set. The RSC satisfies the \textit{Edge-Pair Coverag}\textit{\textcolor{black}{e}}
criterion, because the test set produced by this algorithm satisfies
the \textit{\textcolor{black}{Prime Paths Coverage}}\textcolor{black}{{}
criterion} \cite{offutt2008introduction}. The PPT algorithm is designed
to satisfy $1\leq TDL\leq n$, where $n$ is the length of the longest
path in \textbf{$\mathfrak{\mathcal{G}}$} (excluding the loops) \cite{bures2017prioritized}.
Thus, it satisfies the \textit{Edge-Pair Coverage} criterion, which
is equivalent to $TDL=2$. 

\subsubsection{\label{subsec:Selection-of-the-best-test-set}Selection of the Best
Test Set}

After the selected algorithms have produced the test sets$T_{1}..T_{m}$,
our strategy selects the test set $T$, which has the best value of
the optimality criteria. The test analyst can select the following
options:
\begin{enumerate}
\item \textbf{Selection by single optimality criterion:} A specific optimality
criterion is specified on input. Then, the test set $T$ that has
the best value according to the specified optimality criterion is
selected. The following options are available: $\mid T\mid$, $edges(T)$,
$edges_{h}(T)$, $edges_{m}(T)$, $uedges(T)$, $uedges_{h}(T)$,
$uedges_{m}(T)$, $nodes(T)$, $unodes(T)$, $er(T)$ (the $T$ with
the lowest valu\textcolor{black}{e of these criteria is considered
as optimal) and $e_{h}(T)$, $e_{m}(T)$, $ue_{h}(T)$ and $ue_{m}(T)$
(}the $T$ with the highest valu\textcolor{black}{e of these criteria
is considered as optimal}). However, a test set $T_{1}$ could be
optimal according to one criterion, for instance $\mid T_{1}\mid$,
but be strongly sub-optimal according to another criterion, for instance
$edges(T_{1})$. In such situations, a test set $T_{2}$ with a slightly
higher$\mid T_{2}\mid$ value but whose $edges(T_{2})$ were closer
to the optimum would be a better choice. In these situations, we use
the optimality function explained below.
\item \textbf{Selection by the optimality function:} The optimality function
selects the best test set using several concurrent optimality criteria,
and it is defined as $o(T_{x})=w_{\mid T\mid}(1-\frac{\mid T_{x}\mid}{\frac{\sum_{T\in T_{1}..T_{m}}\mid T\mid}{m}})+w_{edges(T)}(1-\frac{edges(T_{x})}{\frac{\sum_{T\in T_{1}..T_{m}}edges(T)}{m}})+w_{uedges(T)}(1-\frac{uedges(T_{x})}{\frac{\sum_{T\in T_{1}..T_{m}}uedges(T)}{m}})$.
The constants $w_{\mid T\mid}$, $w_{edges(T)}$ and $w_{uedges(T)}$
determine the weight of each specific optimality criterion, $0\leq w_{\mid T\mid}\leq1$,
$0\leq w_{edges(T)}\leq1$, $0\leq w_{uedges(T)}\leq1$ and $w_{\mid T\mid}+w_{edges(T)}+w_{uedges(T)}=1$. 
\item \textbf{Sequence selection:} In this approach, a sequence of optimality
criteria $c_{1}..c_{n}$, is specified on input. When multiple test
sets in $T_{1}..T_{m}$ have the same best value of $c_{1}$, the
best $T$ selection is then based on the $c_{2}$ criterion. If multiple
test sets still have the same best value of $c_{2}$, the selection
of the final $T$ is based on $c_{3}$ and so forth.
\end{enumerate}

\subsubsection{\label{subsec:RSC-Algorithm}RSC Algorithm}

The pseudocode for the Set-Covering Based Solution with Test Set Reduction
(RSC) is specified in Algorithm \ref{alg:Set-Covering-with-Test-Set-Reduction}.

\begin{algorithm*}
\KwIn{$\mathfrak{\mathcal{G}}$, $G$, $PL\in\{high,medium\}$}  
\KwOut{test set $T$}

$P\leftarrow$ test cases satisfying the Prime Path Coverage in $G$ generated by the SC algorithm

$COVER\leftarrow E{}_{h}$ for $PL=high$

$COVER\leftarrow E{}_{h}\cup E_{m}$ for $PL=medium$

$T\leftarrow\emptyset$

\While {$COVER \neq \emptyset$}{
	select $p\in P$ such that $p\cap COVER$ is maximal          

$ADD\leftarrow p\cap COVER$

	$T\leftarrow T+\{p\}$

	$COVER\leftarrow COVER\setminus ADD$
}\tcp{verification of $T$ completeness}
\If {$PL=high$ }{
	$V\leftarrow E{}_{h}$
} \ElseIf{$PL=medium$} { 	$V\leftarrow E{}_{h}\cup E_{m}$ }
\ForEach {$e\in V$}{     \If {$e\notin$ any $t\in T$}{ 		$T$ is invalid 	} }

\caption{\label{alg:Set-Covering-with-Test-Set-Reduction}Set-Covering Based
Solution with Test Set Reduction }
\end{algorithm*}

The principle underlying the RSC is to first employ the SC algorithm
to generate the test set satisfying the Prime Path Coverage (denoted
as $P$). Then, from $P$, the test cases that cover the maximal number
of priority edges that must be present in the test cases are utilized
to build test set $T$ incrementally. 

\section{\label{sec:Experiments}Experiments}

In this section, we describe some experiments performed to demonstrate
the functionality of the proposed test set selection strategy. The
provided data can also be used to compare the test sets produced by
the individual algorithms and when using various optimality criteria.

\subsection{Experimental Method and Set-up}

In the experiments, we execute the algorithms for the following configurations
of the test coverage criteria:
\begin{enumerate}
\item \textbf{\textit{Edge Coverage:}} reduced by $PL=high$ and $PL=medium$.
In this experiment, we compared PPT, RSC, BF, SC and PG. For BF, SC
and PG, the priorities in $\mathfrak{\mathcal{G}}$ were converted
to $R$ by both Atomic and Sequence conversion (see Table \ref{tab:Creation_of_Test_requirements}).
\item \textbf{\textit{Edge-Pair Coverage:}}\textit{ }reduced by $PL=high$
and $PL=medium$. In this experiment, we compared PPT and RSC.
\end{enumerate}
The \textit{Prime Path} \textit{Coverage} criterion was not involved
in the experiments, as its reduction by $PL$ is possible only by
the RSC algorithm. Hence, no alternative algorithm was available for
comparison. The same situation is applicable for $TDL>2$, where the
PPT algorithm is the only option, which reduces the test cases by
$PL$.

Regarding the problem instances, we used 50 SUT models specified by
\textbf{$\mathfrak{\mathcal{G}}$}. To ensure the objective comparability
of all algorithms (and the convertibility of \textbf{$\mathfrak{\mathcal{G}}$}
to $G$ and $R$), the graphs did not contain parallel edges (the
SUTs were modeled so that parallel edges were not needed). The models
were created in the user interface of the Oxygen platform \cite{bures2015pctgen}.
The properties of these models are summarized in Table \ref{tab:Used-problem-instances}.
$\mid E{}_{l}\mid=\mid E\mid-\mid E{}_{h}\mid-\mid E_{m}\mid$. For
the atomic conversion of test requirements (see Table \ref{tab:Creation_of_Test_requirements}),
$R_{ha}$ (resp. $R_{ma}$) denotes a set of test requirements for
$PTL=high$ (resp. $PTL=medium$). For the sequence conversion of
test requirements, $R_{hs}$ (resp. $R_{ms}$) denotes a set of test
requirements for $PTL=high$ (resp. $PTL=medium$). In Table \ref{tab:Used-problem-instances},
we present only$\mid R_{hs}\mid$ and $\mid R_{ms}\mid$, because
$\mid R_{ha}\mid=\mid E_{h}\mid$ and $\mid R_{hm}\mid=\mid E_{h}\mid+\mid E_{m}\mid$.
The number of loops in \textbf{$\mathfrak{\mathcal{G}}$} is denoted
by $loops$. 

\begin{table*}
\scriptsize

\caption{\label{tab:Used-problem-instances}Problem instances used in the experiments}

\centering{}%
\begin{tabular}{|>{\centering}p{1cm}|c|c|c|c|c|c|c||>{\centering}p{1cm}|c|c|c|c|c|c|c|}
\hline 
\textbf{ID} &
$\mid N\mid$ &
$\mid E\mid$ &
$\mid E_{h}\mid$ &
$\mid E{}_{m}\mid$ &
$loops$ &
$\mid R_{hs}\mid$ &
$\mid R_{ms}\mid$ &
\textbf{ID} &
$\mid N\mid$ &
$\mid E\mid$ &
$\mid E_{h}\mid$ &
$\mid E{}_{m}\mid$ &
$loops$ &
$\mid R_{hs}\mid$ &
$\mid R_{ms}\mid$\tabularnewline
\hline 
\hline 
\textbf{1} &
22 &
30 &
8 &
8 &
1 &
6 &
7 &
\textbf{26} &
51 &
67 &
15 &
7 &
0 &
9 &
12\tabularnewline
\hline 
\textbf{2} &
21 &
30 &
8 &
3 &
4 &
5 &
6 &
\textbf{27} &
28 &
39 &
8 &
4 &
1 &
5 &
8\tabularnewline
\hline 
\textbf{3} &
41 &
54 &
10 &
10 &
4 &
9 &
11 &
\textbf{28} &
21 &
22 &
5 &
2 &
0 &
2 &
3\tabularnewline
\hline 
\textbf{4} &
29 &
46 &
11 &
4 &
11 &
9 &
9 &
\textbf{29} &
29 &
37 &
10 &
6 &
0 &
4 &
7\tabularnewline
\hline 
\textbf{5} &
29 &
45 &
17 &
6 &
6 &
15 &
19 &
\textbf{30} &
9  &
11 &
1 &
4 &
0 &
1 &
2\tabularnewline
\hline 
\textbf{6} &
21 &
27 &
9 &
6 &
0 &
8 &
13 &
\textbf{31} &
10  &
13 &
1 &
4 &
0 &
1 &
2\tabularnewline
\hline 
\textbf{7} &
45 &
64 &
18 &
14 &
7 &
15 &
30 &
\textbf{32} &
25 &
27 &
3 &
5 &
0 &
3 &
5\tabularnewline
\hline 
\textbf{8} &
19 &
30 &
10 &
6 &
6 &
9 &
16 &
\textbf{33} &
11 &
15 &
1 &
2 &
0 &
1 &
3\tabularnewline
\hline 
\textbf{9} &
25 &
38 &
11 &
9 &
9 &
8 &
13 &
\textbf{34} &
13 &
19 &
3 &
4 &
0 &
2 &
6\tabularnewline
\hline 
\textbf{10} &
52 &
78 &
9 &
7 &
3 &
6 &
10 &
\textbf{35} &
10  &
15 &
4 &
2 &
4 &
4 &
4\tabularnewline
\hline 
\textbf{11} &
48 &
69 &
10 &
8 &
3 &
4 &
8 &
\textbf{36} &
8 &
10 &
1 &
3 &
3 &
1 &
4\tabularnewline
\hline 
\textbf{12} &
47 &
68 &
9 &
11 &
1 &
5 &
8 &
\textbf{37} &
8 &
11 &
3 &
2 &
3 &
3 &
3\tabularnewline
\hline 
\textbf{13} &
23 &
26 &
9 &
6 &
2 &
5 &
5 &
\textbf{38} &
7  &
12 &
2 &
3 &
5 &
2 &
3\tabularnewline
\hline 
\textbf{14} &
8  &
10 &
1 &
3 &
2 &
1 &
4 &
\textbf{39} &
8  &
11 &
3 &
2 &
2 &
2 &
4\tabularnewline
\hline 
\textbf{15} &
24  &
31 &
10 &
4 &
0 &
2 &
3 &
\textbf{40} &
7  &
9 &
2 &
2 &
0 &
2 &
3\tabularnewline
\hline 
\textbf{16} &
26 &
37 &
10 &
3 &
2 &
3 &
4 &
\textbf{41} &
9  &
11 &
2 &
3 &
0 &
2 &
4\tabularnewline
\hline 
\textbf{17} &
27 &
36 &
6 &
7 &
3 &
6 &
10 &
\textbf{42} &
11 &
14 &
2 &
2 &
2 &
2 &
4\tabularnewline
\hline 
\textbf{18} &
20 &
26 &
1 &
8 &
2 &
1 &
2 &
\textbf{43} &
22 &
27 &
5 &
7 &
0 &
4 &
9\tabularnewline
\hline 
\textbf{19} &
28 &
34 &
1 &
2 &
0 &
1 &
3 &
\textbf{44} &
26  &
38 &
6 &
3 &
3 &
6 &
7\tabularnewline
\hline 
\textbf{20} &
9 &
8 &
3 &
2 &
0 &
3 &
4 &
\textbf{45} &
29  &
45 &
8 &
4 &
4 &
7 &
11\tabularnewline
\hline 
\textbf{21} &
8 &
10 &
2 &
2 &
2 &
2 &
3 &
\textbf{46} &
35 &
48 &
5 &
9 &
4 &
5 &
11\tabularnewline
\hline 
\textbf{22} &
34 &
47 &
13 &
5 &
0 &
7 &
9 &
\textbf{47} &
40 &
54 &
8 &
6 &
0 &
8 &
11\tabularnewline
\hline 
\textbf{23} &
35 &
49 &
8 &
3 &
0 &
7 &
10 &
\textbf{48} &
50 &
74 &
13 &
6 &
6 &
13 &
17\tabularnewline
\hline 
\textbf{24} &
37 &
55 &
16 &
5 &
2 &
9 &
13 &
\textbf{49} &
21 &
27 &
12 &
6 &
0 &
2 &
3\tabularnewline
\hline 
\textbf{25} &
41 &
59 &
15 &
6 &
0 &
10 &
13 &
\textbf{50} &
22 &
23 &
8 &
2 &
0 &
2 &
3\tabularnewline
\hline 
\end{tabular}
\end{table*}

We compared all the options of test set optimality criteria introduced
in Section \ref{subsec:Selection-of-the-best-test-set}: a set of
single optimality criteria, an optimality function and sequence selection.

The test set selection process described in this paper is implemented
as part of the development branch of the Oxygen platform. All the
test set optimality criteria discussed in this paper are calculated
automatically from the produced test cases and provided in a CSV-formatted
report. In the report, the test set $T$ selected by the particular
optimality criteria is also presented, including the algorithm that
generated this test set.

\subsection{\label{subsec:Results-of-the-Experiments}Experimental Results}

\textcolor{black}{In this section, we present a performance comparison
of the PPT, RSC, BF, SC, and PG algorithms using all the test set
optimality criteria discussed in Section \ref{sec:Test-Set-Optimality-Criteria}.
For the comparison that appears in Section \ref{subsec:Comparison-of-individual}
we used the data from Step 3 of Algorithm \ref{alg:The-main-process-Algo}.
}Then, in Section \textcolor{black}{\ref{subsec:Results-of-the-algorithm-selection}},
we provide the results of each specific algorithm selection, which
are the output of Algorithm \textcolor{black}{\ref{alg:The-main-process-Algo}}.

\subsubsection{\label{subsec:Comparison-of-individual}Comparison of Individual
Algorithms}

In Table \ref{tab:Results-Edge-Coverage-PL_high}, we present the
averaged values of the optimality criteria of the test sets produced
for the individual SUT models used in the experiments (introduced
previously in Table \ref{tab:Used-problem-instances}). Table \ref{tab:Results-Edge-Coverage-PL_high}
presents the numbers for \textit{Edge Coverage }and $PL=high$. The
atomic and sequence conversions of test requirements $R$ (see Table
\ref{tab:Creation_of_Test_requirements}) are denoted by ``\textit{atom}''
and ``\textit{seq}'' in the table.

\begin{table*}
\scriptsize

\caption{\label{tab:Results-Edge-Coverage-PL_high}Results of the algorithms
for \textit{Edge Coverage} and $PL=high$}

\centering{}%
\begin{tabular}{|>{\centering}p{3cm}|c|c|c|c|c|c|c|c|}
\hline 
 &
\multicolumn{8}{c|}{Algorithm / $R$ creation method}\tabularnewline
\hline 
Value of optimality criterion - average for all \textbf{$\mathfrak{\mathcal{G}}$}  &
PPT &
RSC &
BF \textit{atom} &
BF \textit{seq} &
SC \textit{atom} &
SC \textit{seq} &
PG \textit{atom} &
PG \textit{seq}\tabularnewline
\hline 
\hline 
$\mid T\mid$  &
2.88 &
3.20 &
5.04 &
4.40 &
5.10 &
4.36 &
4.20 &
4.26\tabularnewline
\hline 
$edges(T)$ &
23.40 &
32.62 &
36.10 &
32.68 &
36.92 &
32.74 &
31.30 &
32.12\tabularnewline
\hline 
$edges_{h}(T)$ &
8.92 &
11.22 &
12.24 &
11.34 &
13.12 &
11.64 &
11.04 &
11.46\tabularnewline
\hline 
$edges_{m}(T)$ &
11.64 &
15.50 &
16.64 &
15.30 &
17.42 &
15.60 &
14.74 &
15.34\tabularnewline
\hline 
$uedges(T)$ &
17.08 &
18.90 &
19.64 &
18.76 &
18.86 &
18.02 &
17.96 &
17.96\tabularnewline
\hline 
$uedges_{h}(T)$ &
7.12 &
7.12 &
7.12 &
7.12 &
7.12 &
7.12 &
7.12 &
7.12\tabularnewline
\hline 
$uedges_{m}(T)$ &
8.86 &
9.28 &
9.30 &
9.20 &
9.22 &
9.18 &
9.14 &
9.16\tabularnewline
\hline 
$nodes(T)$ &
20.52 &
29.42 &
31.06 &
28.28 &
31.82 &
28.38 &
27.10 &
27.86\tabularnewline
\hline 
$unodes(T)$ &
15.52 &
17.28 &
17.88 &
17.06 &
17.18 &
16.40 &
16.32 &
16.36\tabularnewline
\hline 
$er(T)$ &
0.50 &
0.57 &
0.56 &
0.54 &
0.54 &
0.52 &
0.52 &
0.52\tabularnewline
\hline 
$e_{h}(T)$ &
0.41 &
0.38 &
0.36 &
0.39 &
0.38 &
0.40 &
0.40 &
0.40\tabularnewline
\hline 
$e_{m}(T)$ &
0.53 &
0.52 &
0.49 &
0.51 &
0.49 &
0.52 &
0.52 &
0.52\tabularnewline
\hline 
$ue_{h}(T)$ &
0.36 &
0.28 &
0.25 &
0.29 &
0.25 &
0.30 &
0.30 &
0.30\tabularnewline
\hline 
$ue_{m}(T)$ &
0.45 &
0.38 &
0.33 &
0.38 &
0.33 &
0.38 &
0.39 &
0.39\tabularnewline
\hline 
\end{tabular}
\end{table*}

\begin{figure*}
\begin{centering}
\includegraphics[width=12cm]{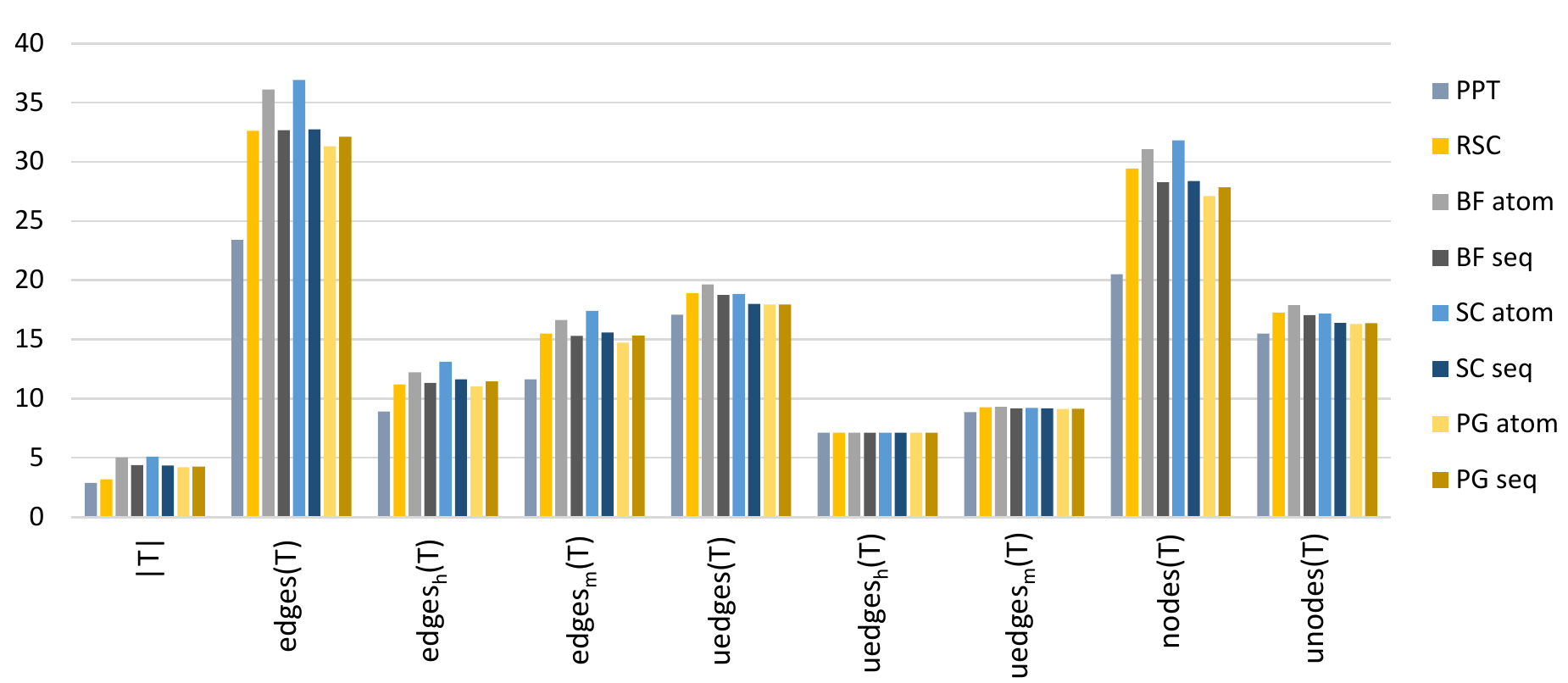}
\par\end{centering}
\caption{\label{fig:Comparison-of-the-algorithms-EdgeCoverage-PL_high}Algorithm
comparison for \textit{Edge Coverage} and $PL=high$}
\end{figure*}

\begin{figure}
\begin{centering}
\includegraphics[width=8cm]{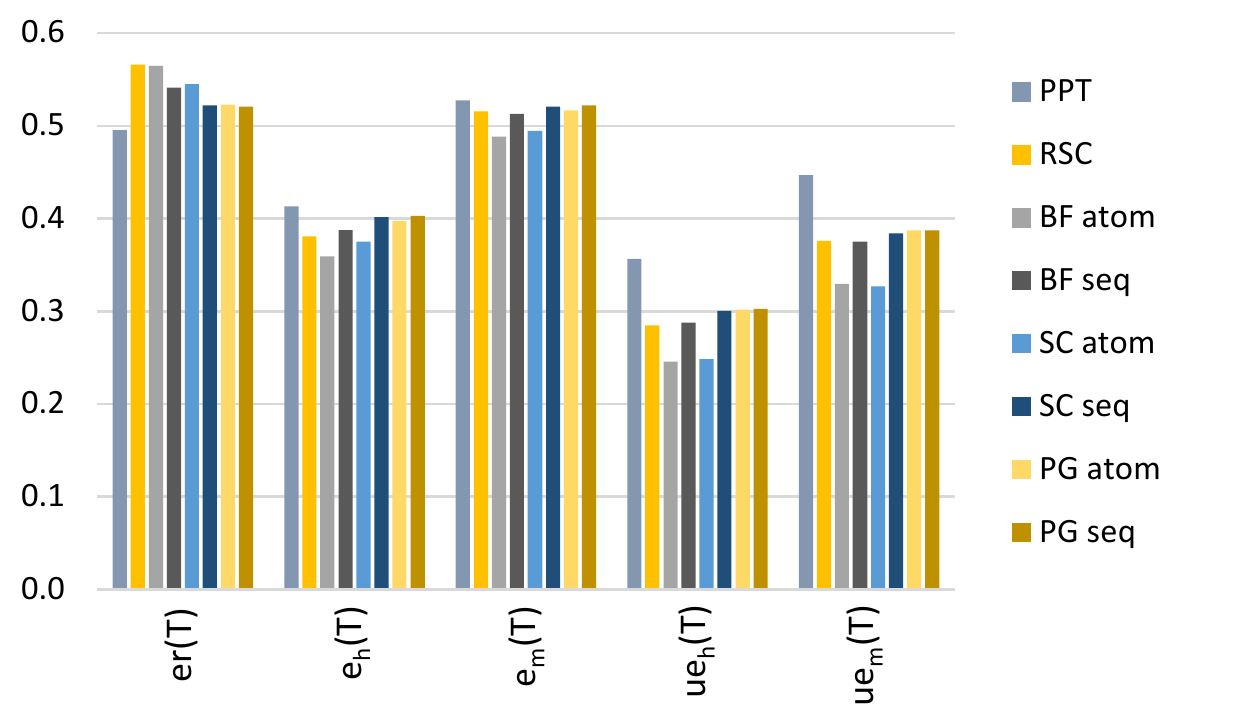}
\par\end{centering}
\caption{\label{fig:Comparison-of-the-algorithms-EdgeCoverage-PL_high-2ndPart}Algorithm
comparison for \textit{Edge Coverage} and $PL=high$}
\end{figure}

Figures \ref{fig:Comparison-of-the-algorithms-EdgeCoverage-PL_high}
and \ref{fig:Comparison-of-the-algorithms-EdgeCoverage-PL_high-2ndPart}
provide a visual comparison, using these averaged values to compare
the individual algorithms.

Table \ref{tab:Results-Edge-Coverage-PL_medium} lists the averaged
values of the optimality criteria of the test sets produced for the
individual SUT models for the \textit{Edge Coverage }and $PL=medium$
criterion and $PL=medium$.

\begin{table*}
\scriptsize

\caption{\label{tab:Results-Edge-Coverage-PL_medium}Results of the algorithms
for \textit{Edge Coverage} and $PL=medium$}

\centering{}%
\begin{tabular}{|>{\centering}p{3cm}|c|c|c|c|c|c|c|c|}
\hline 
 &
\multicolumn{8}{c|}{Algorithm / $R$ creation method}\tabularnewline
\hline 
Value of optimality criterion - average for all \textbf{$\mathfrak{\mathcal{G}}$}  &
PPT &
RSC &
BF \textit{atom} &
BF \textit{seq} &
SC \textit{atom} &
SC \textit{seq} &
PG \textit{atom} &
PG \textit{seq}\tabularnewline
\hline 
\hline 
$\mid T\mid$  &
4.38 &
4.76 &
7.66 &
6.80 &
7.60 &
6.52 &
5.96 &
6.20\tabularnewline
\hline 
$edges(T)$ &
35.34 &
46.04 &
54.18 &
50.46 &
54.20 &
48.94 &
44.84 &
47.28\tabularnewline
\hline 
$edges_{h}(T)$ &
10.56 &
13.26 &
15.40 &
15.06 &
16.32 &
15.30 &
13.48 &
14.86\tabularnewline
\hline 
$edges_{m}(T)$ &
17.22  &
21.54 &
24.44 &
24.24 &
25.38 &
24.28 &
21.38 &
23.48\tabularnewline
\hline 
$uedges(T)$ &
22.60 &
24.06 &
24.72 &
23.60 &
23.98 &
22.76 &
22.80 &
22.68\tabularnewline
\hline 
$uedges_{h}(T)$ &
7.12 &
7.12 &
7.12 &
7.12 &
7.12 &
7.12 &
7.12 &
7.12\tabularnewline
\hline 
$uedges_{m}(T)$ &
12.08 &
12.08 &
12.08 &
12.08 &
12.08 &
12.08 &
12.08 &
12.08\tabularnewline
\hline 
$nodes(T)$ &
30.96 &
41.28 &
46.52 &
43.66 &
46.60 &
42.42 &
38.88 &
41.08\tabularnewline
\hline 
$unodes(T)$ &
20.80 &
22.28 &
22.84 &
21.80 &
22.16 &
21.04 &
21.06 &
20.96\tabularnewline
\hline 
$er(T)$ &
0.69 &
0.74 &
0.75 &
0.72 &
0.74 &
0.70 &
0.70 &
0.69\tabularnewline
\hline 
$e_{h}(T)$ &
0.30 &
0.29 &
0.28 &
0.31 &
0.29 &
0.32 &
0.31 &
0.32\tabularnewline
\hline 
$e_{m}(T)$ &
0.53 &
0.51 &
0.47 &
0.53 &
0.48 &
0.54 &
0.52 &
0.54\tabularnewline
\hline 
$ue_{h}(T)$ &
0.22 &
0.18 &
0.14 &
0.16 &
0.14 &
0.17 &
0.18 &
0.17\tabularnewline
\hline 
$ue_{m}(T)$ &
0.42 &
0.33 &
0.26 &
0.31 &
0.25 &
0.33 &
0.34 &
0.33\tabularnewline
\hline 
\end{tabular}
\end{table*}

\begin{figure*}
\begin{centering}
\includegraphics[width=12cm]{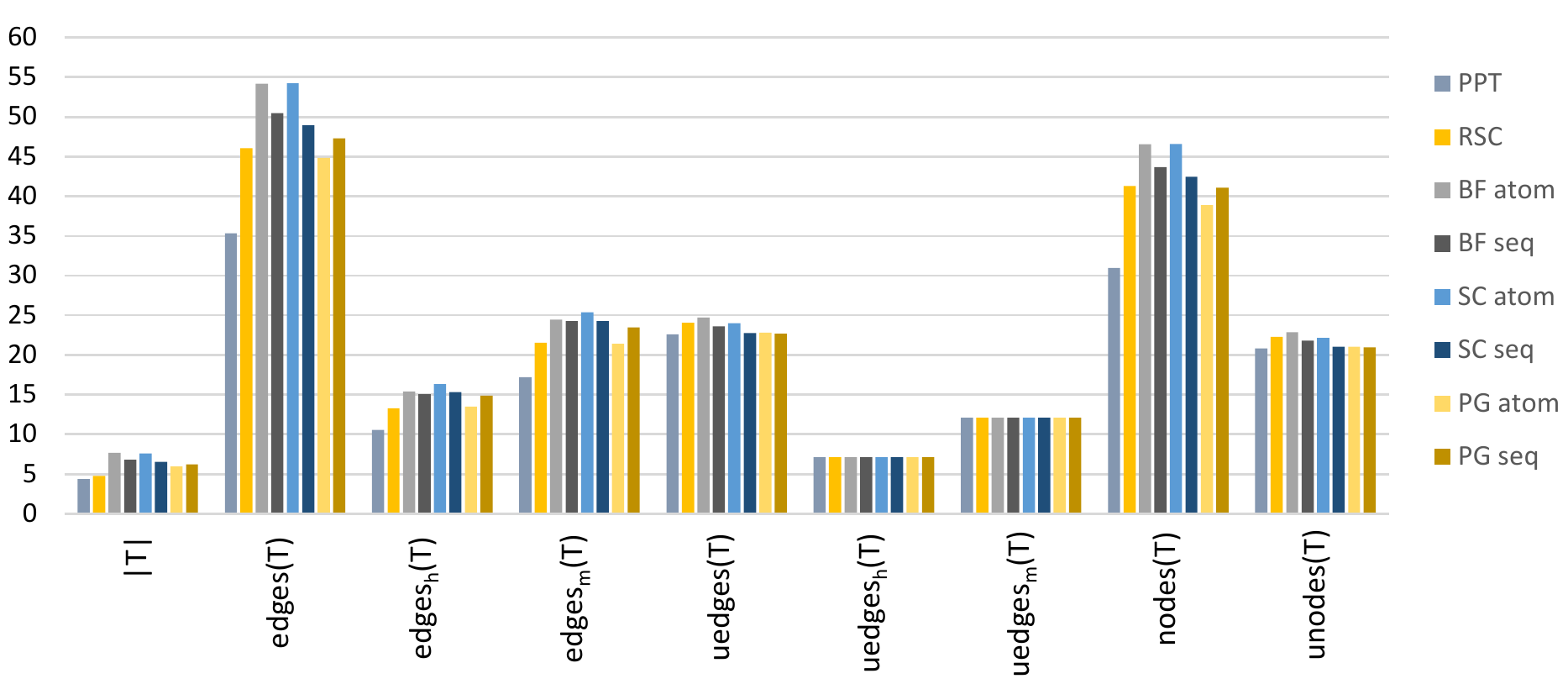}
\par\end{centering}
\caption{\label{fig:Comparison-of-the-algorithms-EdgeCoverage-PL_medium}Algorithm
comparison for \textit{Edge Coverage} and $PL=medium$}
\end{figure*}

\begin{figure}
\begin{centering}
\includegraphics[width=8cm]{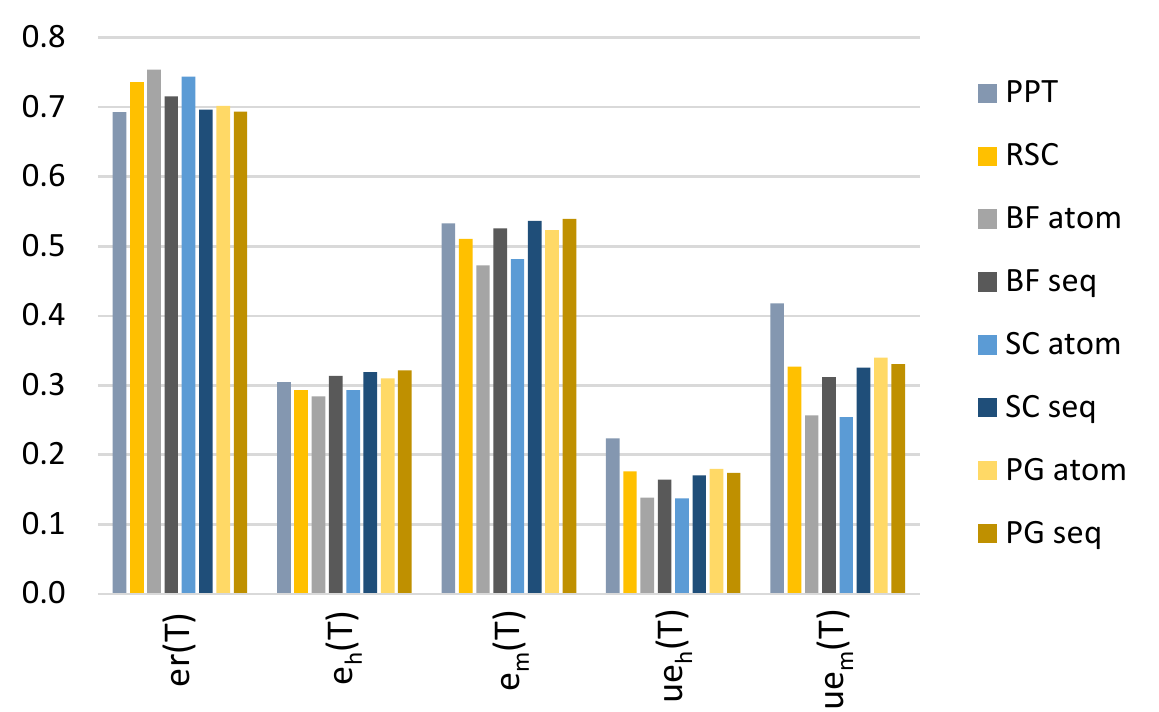}
\par\end{centering}
\caption{\label{fig:Comparison-of-the-algorithms-EdgeCoverage-PL_medium--2ndPart}Algorithm
comparison for \textit{Edge Coverage} and $PL=medium$}
\end{figure}

To better compare the algorithms using the values presented in \ref{tab:Results-Edge-Coverage-PL_medium},
Figures \ref{fig:Comparison-of-the-algorithms-EdgeCoverage-PL_medium}
and \ref{fig:Comparison-of-the-algorithms-EdgeCoverage-PL_medium--2ndPart}
shows a graphical summary.

Table \ref{tab:Results-Edge-Pair-Coverage} shows a comparison of
the PPT and RSC algorithms for the \textit{Edge-Pair Coverage} criterion
and $PL\in\{high,medium\}$. Here, the only relevant algorithms are
PPT and RSC, because these algorithms allow test set reduction by
PL and can satisfy the \textit{Edge-Pair Coverage} criterion before
the test set reduction by PL.

\begin{table}
\scriptsize

\caption{\label{tab:Results-Edge-Pair-Coverage}Results of the algorithms for
Edge-Pair Coverage}

\centering{}%
\begin{tabular}{|>{\centering}p{2cm}|c|c|c|c|}
\hline 
 &
\multicolumn{2}{c|}{$PL=high$} &
\multicolumn{2}{c|}{ $PL=medium$}\tabularnewline
\hline 
Value of optimality criterion - average for all \textbf{$\mathfrak{\mathcal{G}}$}  &
PPT &
RSC &
PPT &
RSC\tabularnewline
\hline 
\hline 
$\mid T\mid$  &
5.28 &
3.20 &
7.78 &
4.76\tabularnewline
\hline 
$edges(T)$ &
43.74 &
32.62 &
62.88 &
46.04\tabularnewline
\hline 
$edges_{h}(T)$ &
15.50 &
11.22 &
18.32 &
13.26\tabularnewline
\hline 
$edges_{m}(T)$ &
20.72 &
15.50 &
29.18 &
21.54\tabularnewline
\hline 
$uedges(T)$ &
23.04 &
18.90 &
28.00 &
24.06\tabularnewline
\hline 
$uedges_{h}(T)$ &
7.12 &
7.12 &
7.12 &
7.12\tabularnewline
\hline 
$uedges_{m}(T)$ &
9.92 &
9.28 &
12.08 &
12.08\tabularnewline
\hline 
$nodes(T)$ &
38.46 &
29.42 &
55.10 &
41.28\tabularnewline
\hline 
$unodes(T)$ &
21.46 &
17.28 &
26.30 &
22.28\tabularnewline
\hline 
$er(T)$ &
0.65 &
0.57 &
0.86 &
0.74\tabularnewline
\hline 
$e_{h}(T)$ &
0.36 &
0.38 &
0.27 &
0.29\tabularnewline
\hline 
$e_{m}(T)$ &
0.49 &
0.52 &
0.47 &
0.51\tabularnewline
\hline 
$ue_{h}(T)$ &
0.20 &
0.28 &
0.12 &
0.18\tabularnewline
\hline 
$ue_{m}(T)$ &
0.29 &
0.38 &
0.21 &
0.33\tabularnewline
\hline 
\end{tabular}
\end{table}

\begin{figure*}
\begin{centering}
\includegraphics[width=10cm]{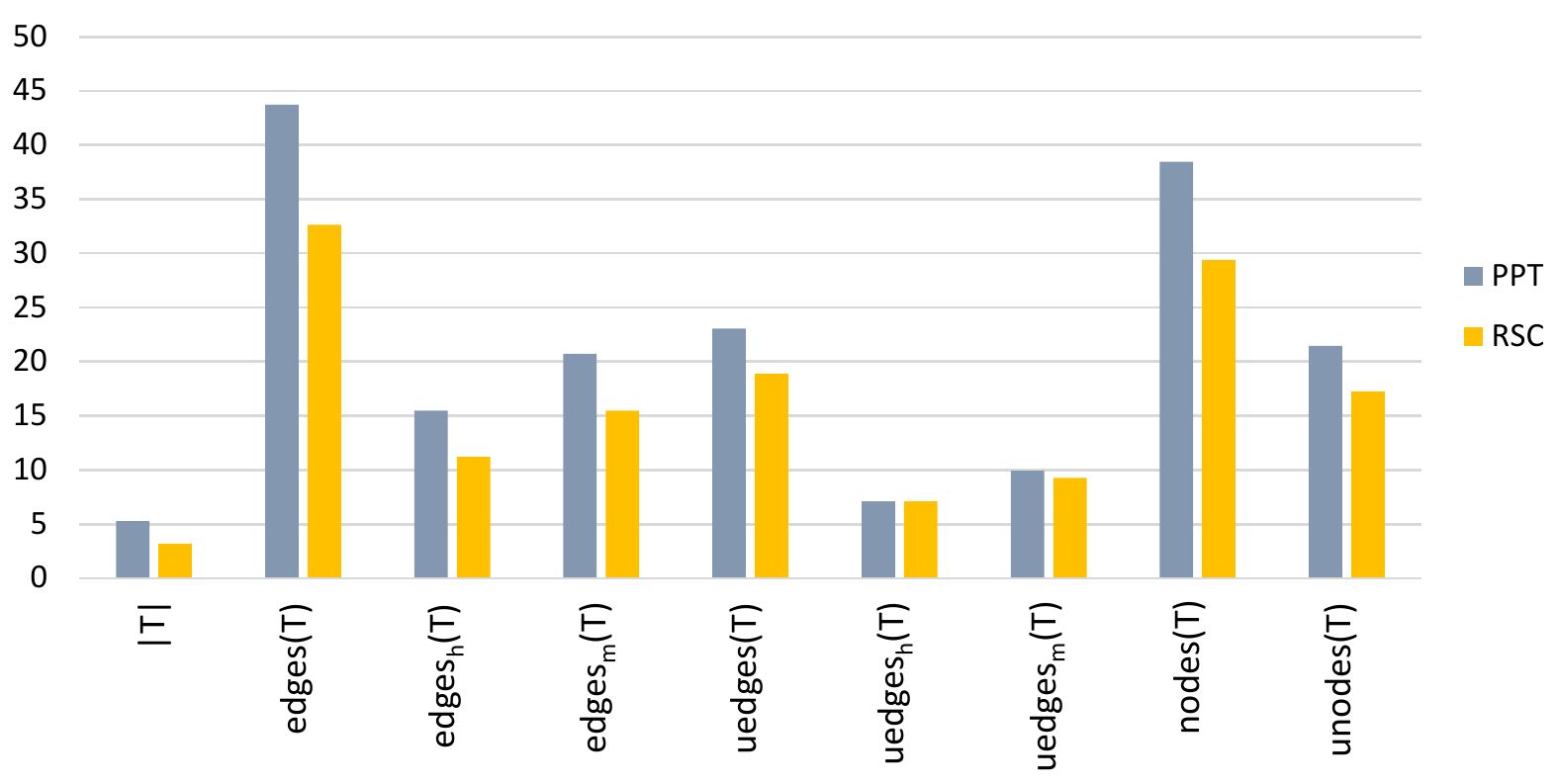}
\par\end{centering}
\caption{\label{fig:Comparison-of-the-algorithms-Edge_PAIR_Coverage-PL_high}Algorithm
comparison for \textit{Edge-Pair Coverage} and $PL=high$}
\end{figure*}

\begin{figure}
\begin{centering}
\includegraphics[width=6cm]{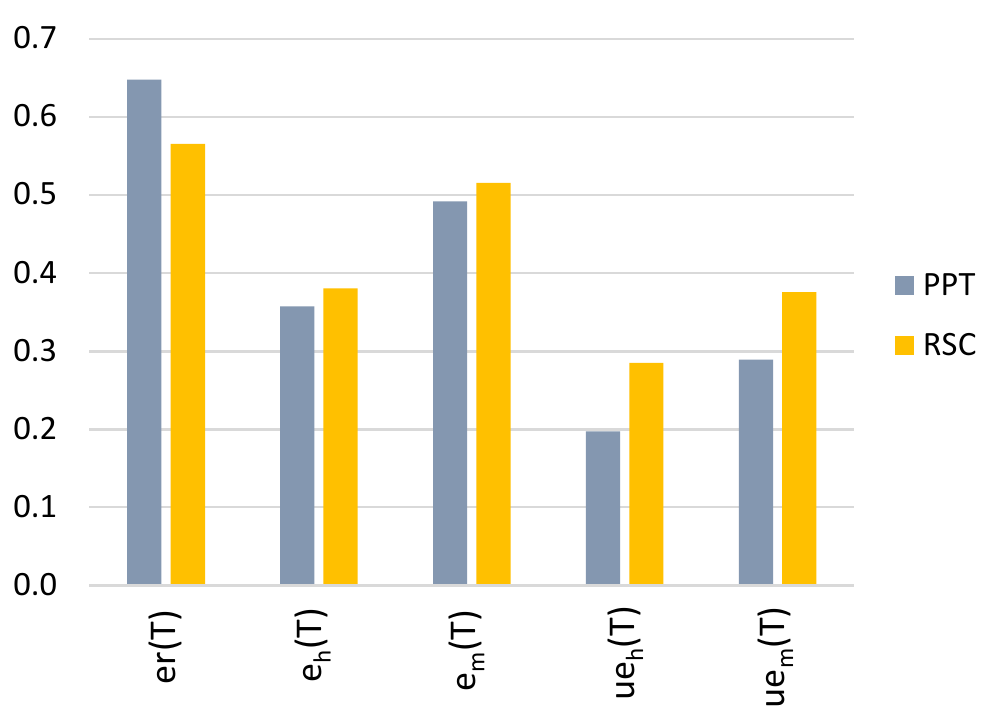}
\par\end{centering}
\caption{\label{fig:Comparison-of-the-algorithms-Edge_PAIR_Coverage-PL_high-2ndPart}Algorithm
comparison for \textit{Edge-Pair Coverage} and $PL=high$}
\end{figure}

A comparison of the individual algorithms for the \textit{Edge-Pair
Coverage }criterion is depicted in Figures \ref{fig:Comparison-of-the-algorithms-Edge_PAIR_Coverage-PL_high}
and \ref{fig:Comparison-of-the-algorithms-Edge_PAIR_Coverage-PL_high-2ndPart}
for $PL=high$. 

\begin{figure*}
\begin{centering}
\includegraphics[width=10cm]{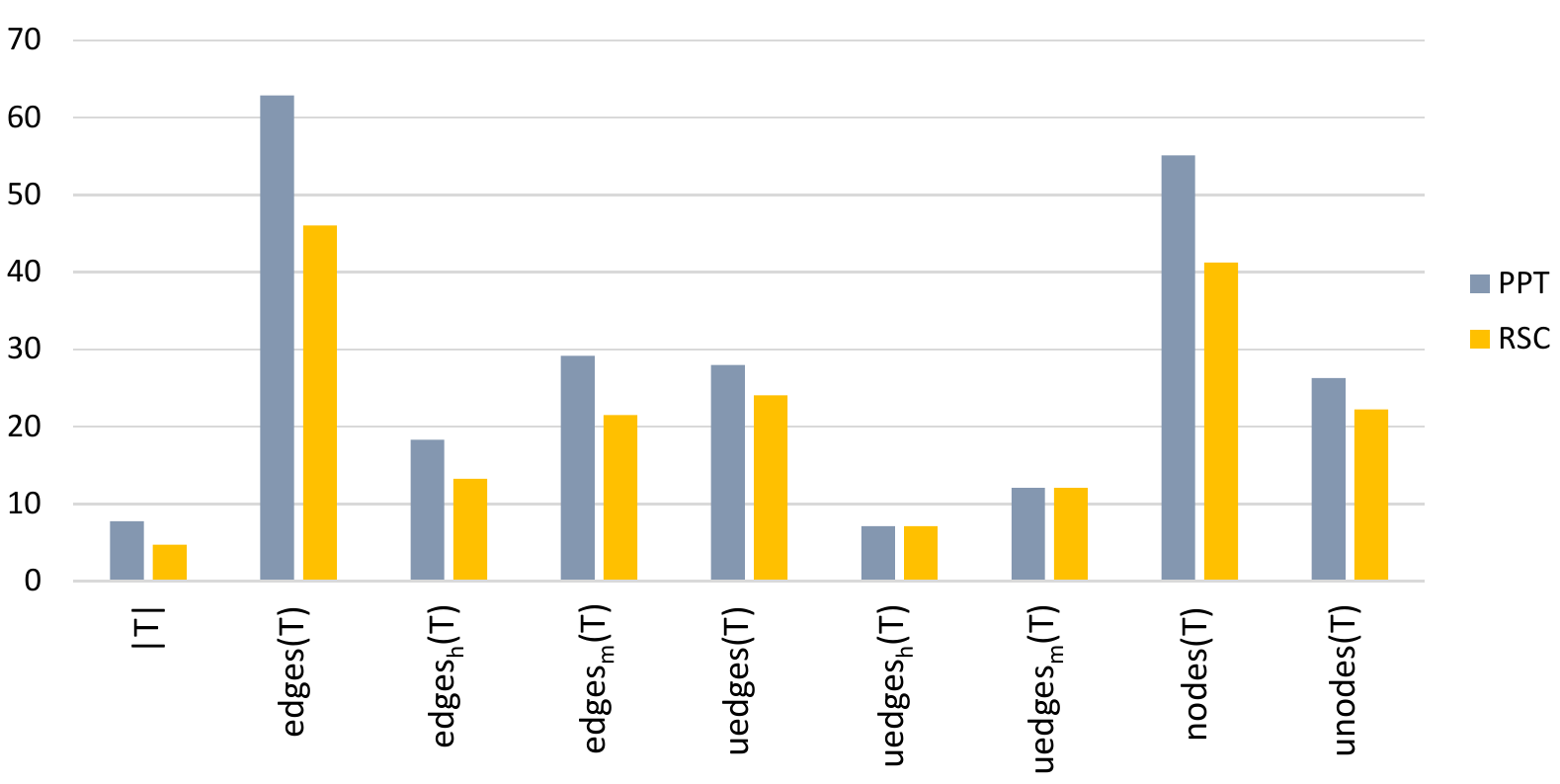}
\par\end{centering}
\caption{\label{fig:Comparison-of-the-algorithms-Edge_PAIR_Coverage-PL_medium}Algorithm
comparison for \textit{Edge-Pair Coverage} and $PL=medium$}
\end{figure*}

\begin{figure}
\begin{centering}
\includegraphics[width=6cm]{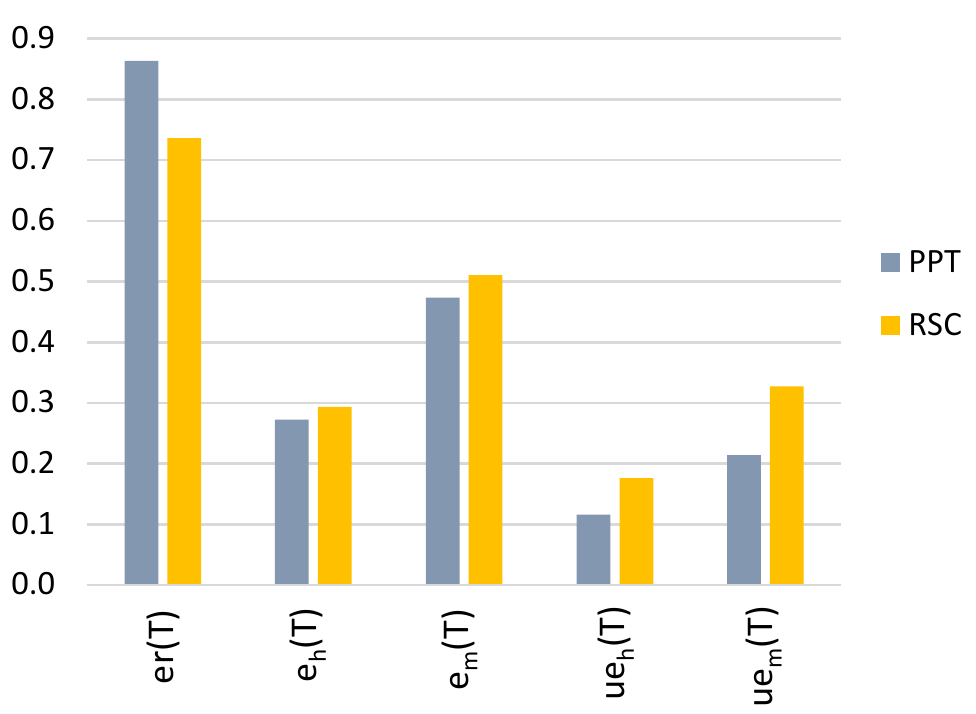}
\par\end{centering}
\caption{\label{fig:Comparison-of-the-algorithms-Edge_PAIR_Coverage-PL_medium-2ndPart}Algorithm
comparison for \textit{Edge-Pair Coverage} and $PL=medium$}
\end{figure}

Figures \ref{fig:Comparison-of-the-algorithms-Edge_PAIR_Coverage-PL_medium}
and \ref{fig:Comparison-of-the-algorithms-Edge_PAIR_Coverage-PL_medium-2ndPart}
depict this comparison for $PL=medium$.

We analyze the data in Section \ref{sec:Discussion}.

\subsubsection{\label{subsec:Results-of-the-algorithm-selection}Algorithm Selection
Results}

In this section, we present the results related to the functionality
of the proposed test set selection strategy. We start with detailed
data that demonstrate the test set selection strategy. Table \ref{tab:Results-of-the-TS-selection-mechanism--edge-coverage--PTL-high}
summarizes the execution of the test set selection strategy for the
\textit{Edge Coverage }criterion and $PL=high$. For each of the problem
instances and optimality criteria, we list the algorithm that produced
the optimal $T$ based on the selected criteria. \textcolor{black}{Table
\ref{tab:Results-of-the-TS-selection-mechanism--edge-coverage--PTL-high}
presents the results of the first twelve selected problem instances
as an example. }

\textbf{OPT} denotes an optimality function (refer to Section \ref{subsec:Selection-of-the-best-test-set}).
In all the experiments described in this paper, the optimality function
was configured with parameters $w_{\mid T\mid}=0.3$, $w_{edges(T)}=0.4$
and $w_{uedges(T)}=0.3$.

\textbf{SEQ} denotes sequence selection (refer to Section \ref{subsec:Selection-of-the-best-test-set}).
In all of the experiments, the following sequence of optimality criteria
was adopted: $\mid T\mid$, $edges(T)$, \textcolor{black}{$uedges(T)$.
The selection sequence starts with $\mid T\mid$.}

In Table \ref{tab:Results-of-the-TS-selection-mechanism--edge-coverage--PTL-high}
we present the name of the algorithm (or algorithms) that produced
the optimal test set based on the particular criterion. For simple
optimality criteria, the algorithm name is followed by the value of
the optimality criterion (in brackets). When multiple algorithms can
provide an optimal $T$ for a particular criterion, more algorithms
are listed. Cases in which more than three algorithms provided an
optimal $T$ for a particular criterion are denoted as $n(M)$, where
$n$ denotes the number of algorithms and $M$ denotes the value of
the optimality criterion.

Atomic and sequence conversions of test requirements $R$ (refer to
Table \ref{tab:Creation_of_Test_requirements}) are denoted \textcolor{black}{by
``}\textit{\textcolor{black}{a}}\textcolor{black}{'' and ``}\textit{\textcolor{black}{s}}\textcolor{black}{''
}postfixes, respectively, in italics following the name of the algorithm.
ID denotes the ID of the problem instance $\mathfrak{\mathcal{G}}$.

\begin{sidewaystable*}
\scriptsize
\begin{centering}
\caption{\label{tab:Results-of-the-TS-selection-mechanism--edge-coverage--PTL-high}Results
of the test set selection strategy for the individual problem instances
for \textit{Edge Coverage} and $PL=high$}
\par\end{centering}
\centering{}%
\begin{tabular}{|>{\centering}p{0.6cm}|>{\centering}p{1.5cm}|>{\centering}p{1.5cm}|>{\centering}p{1.5cm}|>{\centering}p{1.5cm}|>{\centering}p{1.5cm}|>{\centering}p{1.5cm}|>{\centering}p{1.7cm}|>{\centering}p{1.7cm}|>{\centering}p{1.7cm}|>{\centering}p{1.5cm}|>{\centering}p{1.5cm}|}
\hline 
 &
\multicolumn{11}{c|}{\textbf{Optimality criterion}}\tabularnewline
\hline 
\textbf{ID} &
$\mid T\mid$ &
$edges(T)$ &
$edges_{h}(T)$ &
$uedges(T)$ &
$nodes(T)$ &
$unodes(T)$ &
$er(T)$ &
$e_{h}(T)$ &
$e_{m}(T)$ &
\textbf{OPT} &
\textbf{SEQ}\tabularnewline
\hline 
\hline 
1 &
RSC(4) PPT(4) &
PPT(18) &
RSC(8) PPT(8) &
PPT(17) &
PPT(14) &
PPT(14) &
PPT(0.57) &
\textit{4}(0.46) &
RSC(0.90) &
PPT &
PPT\tabularnewline
\hline 
2 &
RSC(3) PPT(3) &
PPT(22) &
PG(8) PPT(8) &
PPT(22) &
PPT(19) &
PPT(19) &
PPT(0.73) &
PPT(0.36) &
PPT(0.50) &
PPT &
PPT\tabularnewline
\hline 
3 &
RSC(6) PPT(6) &
SC\textit{a}(67) &
SC\textit{a}(13) &
\textit{4}(35) &
SC\textit{a}(60) &
\textit{4}(34) &
\textit{4}(0.65) &
RSC(0.22) &
PG\textit{a}(0.41) &
SC\textit{a} &
PPT\tabularnewline
\hline 
4 &
PPT(2) &
PPT(31) &
PPT(13) &
PPT(26) &
PPT(29) &
PPT(25) &
PPT(0.56) &
PPT(0.42) &
PPT(0.55) &
PPT &
PPT\tabularnewline
\hline 
5 &
PPT(6) &
PPT(35) &
PPT(19) &
PG\textit{a}(30)

PPT(30) &
PPT(29) &
PG\textit{a}(26)

PPT(26) &
PG\textit{a}(0.67)

PPT(0.67) &
PPT(0.54) &
PPT(0.66) &
PPT &
PPT\tabularnewline
\hline 
6 &
RSC(4) PPT(4) &
RSC(28) PPT(28) &
RSC(12) PPT(12) &
\textit{4}(19) &
RSC(24) PPT(24) &
\textit{4}(17) &
\textit{4}(0.70) &
RSC(0.43) PPT(0.43) &
BF\textit{a}(0.41) &
RSC &
RSC

PPT\tabularnewline
\hline 
7 &
RSC(11) PPT(11) &
PPT(99) &
PPT(26) &
PPT(45) &
PPT(88) &
PG\textit{a}(42)

PG\textit{s}(42)

SC\textit{s}(42) &
PPT(0.70) &
PPT(0.26) &
PPT(0.60) &
PPT &
PPT\tabularnewline
\hline 
8 &
PPT(3) &
PPT(37) &
PPT(14) &
PPT(21) &
PPT(34) &
PPT(20) &
PPT(0.70) &
BF\textit{s}(0.43) &
RSC(0.56) &
PPT &
PPT\tabularnewline
\hline 
9 &
PPT(2) &
PPT(31) &
PPT(15) &
PPT(22) &
PPT(29) &
PPT(21) &
PPT(0.58) &
PPT(0.48) &
PPT(0.65) &
PPT &
PPT\tabularnewline
\hline 
10 &
RSC(3) PPT(3) &
RSC(26) PPT(26) &
RSC(10) PPT(10) &
RSC(25) PPT(25) &
RSC(23) PPT(23) &
RSC(22) PPT(22) &
RSC(0.32) PPT(0.32) &
PG\textit{s}(0.44)

SC\textit{s}(0.44) &
PG\textit{s}(0.51)

SC\textit{s}(0.51) &
RSC

PPT &
RSC

PPT\tabularnewline
\hline 
11 &
RSC(3) PPT(3) &
RSC(22) PPT(22) &
PPT(12) &
RSC(19) &
RSC(19) PPT(19) &
RSC(16) &
RSC(0.28) &
RSC(0.59) &
RSC(0.59) PPT(0.59) &
RSC &
RSC\tabularnewline
\hline 
12 &
RSC(3) PPT(3) &
RSC(27) &
RSC(12) PPT(12) &
\textit{4}(22) &
RSC(24) &
\textit{4}(20) &
\textit{4}(0.32) &
RSC(0.44) &
PG\textit{a}(0.63)

PG\textit{s}(0.63)

SC\textit{s}(0.63) &
RSC &
RSC\tabularnewline
\hline 
\end{tabular}
\end{sidewaystable*}

\begin{figure*}
\begin{centering}
\includegraphics[width=10cm]{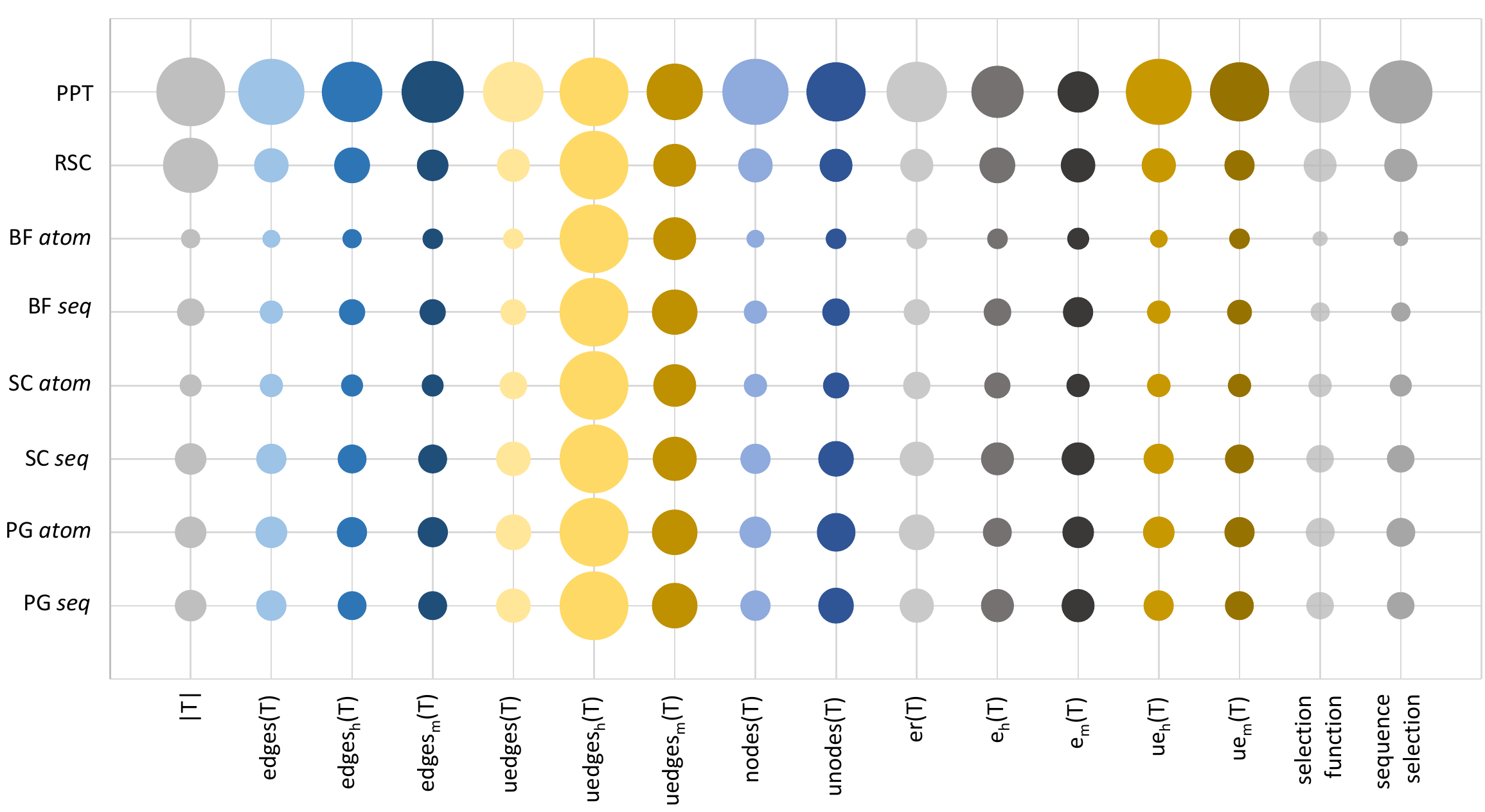}
\par\end{centering}
\caption{\label{fig:Overal-statistics-of-the-TS-selection-mechanism--edge-coverage--PTL-high}Overall
statistics of the test set selection strategy for \textit{Edge Coverage}
and $PL=high$}
\end{figure*}

Figure \ref{fig:Overal-statistics-of-the-TS-selection-mechanism--edge-coverage--PTL-high}
shows the overall statistics for the test set selection strategies
for the \textit{Edge Coverage} criterion and $PL=high$. The x-axis
reflects the optimality criteria, and the y-axis presents the individual
algorithms. The bubble size represents the number of problem instances
for which an algorithm produced an optimal $T$ using a specific criterion.\textcolor{black}{{}
The maximum bubble size is 50 (the number of problem instances). For
the case of $uedges_{h}(T)$, all the algorithms achieved the maximum
value}, as $uedges_{h}(T)=\mid E_{h}\mid$ for $T$ in all the cases.

\begin{figure*}
\begin{centering}
\includegraphics[width=10cm]{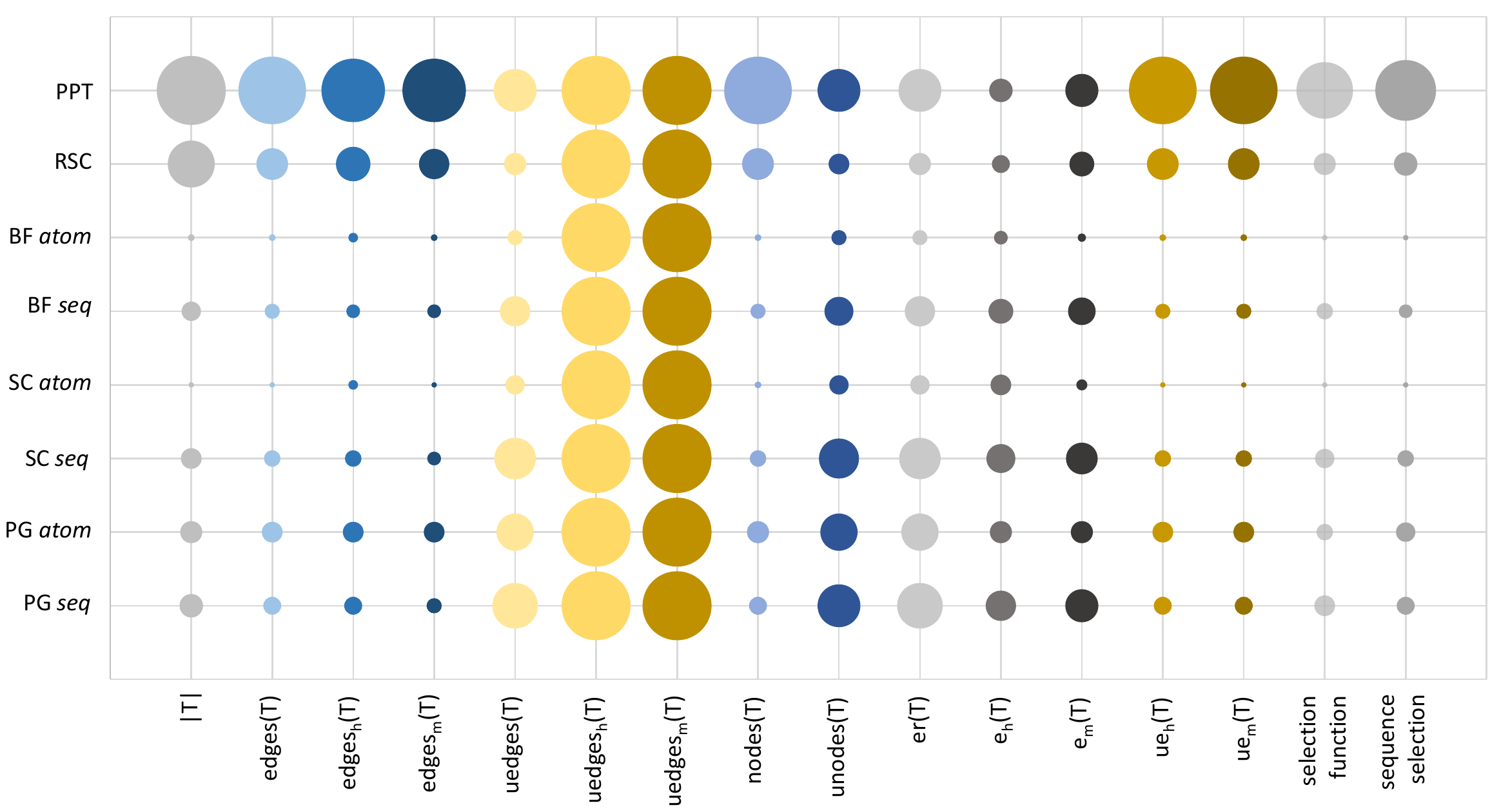}
\par\end{centering}
\caption{\label{fig:Overal-statistics-of-the-TS-selection-mechanism--edge-coverage--PTL-medium}Overall
statistics of the test set selection strategy for \textit{Edge Coverage}
and $PL=medium$}
\end{figure*}

Using the same schema, Figure \textcolor{black}{\ref{fig:Overal-statistics-of-the-TS-selection-mechanism--edge-coverage--PTL-medium}}
shows the overall statistical results of the test set selection strategies
for the \textit{Edge Coverage} criterion and $PL=medium$. \textcolor{black}{In
this case, all the employed algorithms provided a} $T$ having $uedges_{h}(T)=\mid E_{h}\mid$
and $uedges_{m}(T)=\mid E_{h}\mid+\mid E_{m}\mid$.

\begin{figure*}
\begin{centering}
\includegraphics[width=10cm]{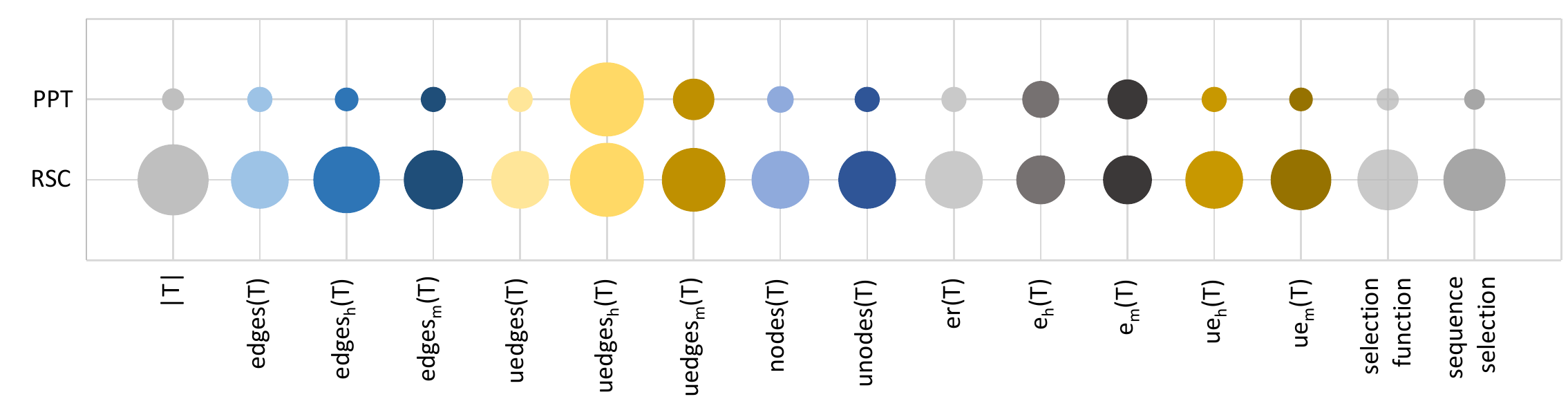}
\par\end{centering}
\caption{\label{fig:Overal-statistics-of-the-TS-selection-mechanism--edge-PAIR-coverage--PTL-high}Overall
statistics of the test set selection strategy for \textit{Edge-Pair
Coverage} and $PL=high$}
\end{figure*}

\begin{figure*}
\begin{centering}
\includegraphics[width=10cm]{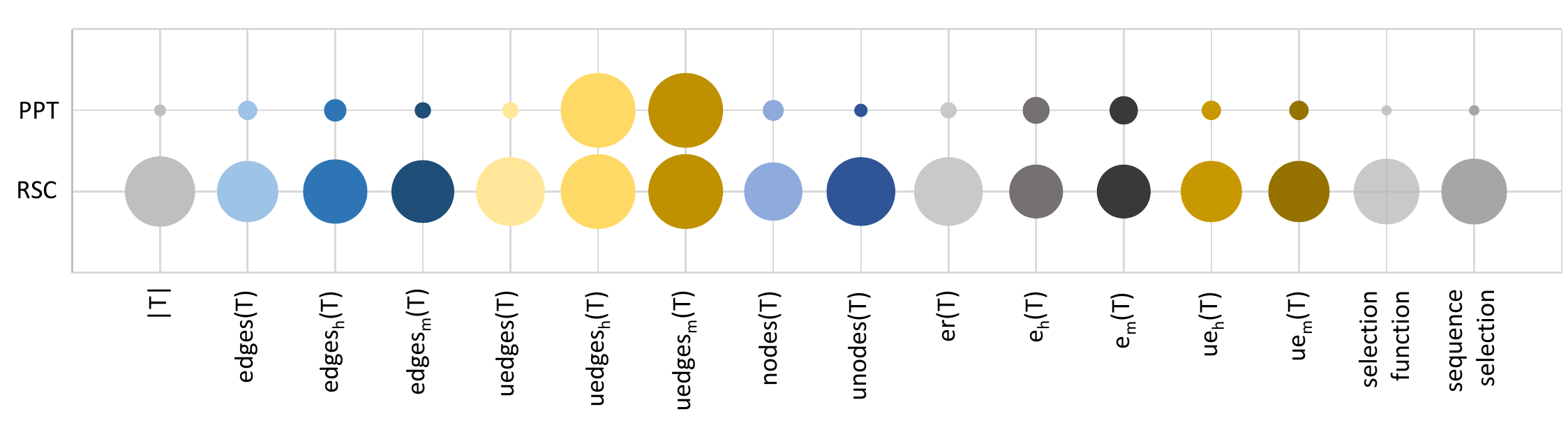}
\par\end{centering}
\caption{\label{fig:Overal-statistics-of-the-TS-selection-mechanism--edge-PAIR-coverage--PTL-medium}Overall
statistics of the test set selection strategy for \textit{Edge-Pair
Coverage} and $PL=medium$}
\end{figure*}

The same system is used in Figure \ref{fig:Overal-statistics-of-the-TS-selection-mechanism--edge-PAIR-coverage--PTL-high},
which depicts the overall statistics of the test set selection strategies
for \textit{Edge-Pair Coverage} and $PL=high$, and in Figure \ref{fig:Overal-statistics-of-the-TS-selection-mechanism--edge-PAIR-coverage--PTL-medium},
which presents the statistics for \textit{Edge-Pair Coverage} and
$PL=medium$. 

\section{\label{sec:Discussion}Discussion}

From the data presented in Section \ref{subsec:Results-of-the-Experiments},
several conclusions can be made. 

Starting with a \textbf{comparison of the algorithms} using the average
values of optimality criteria computed for 50 different problem instances
(Table \ref{tab:Used-problem-instances}), the results differ significantly
based on the test coverage level. For \textit{Edge Coverage}, the
PPT algorithm provides the best results in terms of average statistics.
The difference between the average value for PPT and the other algorithms
is the most significant for the optimality criteria$\mid T\mid$,
$edges(T)$ and $nodes(T)$. This result can be observed for both
$PL=high$ (Table \ref{tab:Results-Edge-Coverage-PL_high}, Figure
\ref{fig:Comparison-of-the-algorithms-EdgeCoverage-PL_high}) and
$PL=medium$ (Table \ref{tab:Results-Edge-Coverage-PL_medium}, Figure
\ref{fig:Comparison-of-the-algorithms-EdgeCoverage-PL_medium}). For
$PL=high$, the difference in $\mid T\mid$ between PPT and RSC is
10\%, and it is greater than 31\% between PPT and each of the BF,
SC and PG algorithms. The difference in $edges(T)$ between PPT and
all the other algorithms is greater than 25\%. For $nodes(T)$, this
difference is greater than 24\%. For $PL=medium$, the differences
are slightly lower in general. For $\mid T\mid$, the difference between
PPT and RSC is 8\%, and it is greater than 27\% between PPT and each
of the BF, SC and PG algorithms. The difference between PPT and all
the other algorithms is greater than 21\% for $edges(T)$ and greater
than 20\% for $nodes(T)$.

Additionally, for the optimality criteria based on unique priority
edges, $ue_{h}(T)$ and $ue_{m}(T)$, the average value of optimality
criteria differs significantly for the PPT algorithm. Higher values
of $ue_{h}(T)$ and $ue_{m}(T)$ result in test sets closer to optimum.
For $PL=high$, PPT is higher than all the other algorithms by 18\%
for $ue_{h}(T)$ and 15\% for $ue_{m}(T)$.

For the rest of the optimality criteria, the differences are not as
significant; however, similar results are still present in the data.

Generally, the RSC algorithm yields results relatively similar to
the BF, PG and SC algorithms; however, exceptions can be found. For
$PL=high$ (Table \ref{tab:Results-Edge-Coverage-PL_high}, Figure
\ref{fig:Comparison-of-the-algorithms-EdgeCoverage-PL_high}), the
RSC algorithm is outperformed by the PG and SC algorithms for $uedges(T)$,
$nodes(T)$, $unodes(T)$, and $er(T)$. At this priority level, the
RSC algorithm does not outperform any of other algorithms.

The situation changes when $PL=medium$ (Table \ref{tab:Results-Edge-Coverage-PL_medium},
Figure \ref{fig:Comparison-of-the-algorithms-EdgeCoverage-PL_medium}),
which, in practical terms, means that the algorithms process more
priority edges. From the data, RSC exhibits better performance in
this case. It is outperformed by the PG and SC algorithms only for
the $uedges(T)$ and $unodes(T)$ optimality criteria. In contrast,
the RSC outperforms BF, PG and SC for $edges(T)$ and $edges_{h}(T)$,
which can be considered as important criteria of test set optimality.

For the total test cases $\mid T\mid$, the RSC yields the better
results than do the BF, PG and SC for both $PL=high$ and $PL=medium$.
However, $\mid T\mid$ itself as an indicator of test set optimality
is probably insufficient; the total number of test steps (e.g. $edges(T)$
or $nodes(T)$) are more reliable metrics. 

Some other conclusions can be drawn from the data for the \textit{Edge
Coverage }criterion. For $PL=high$, the differences in the results
for the\textcolor{black}{{} atomic and sequence conversion of the test
requirements} $R$ (refer to Table \ref{tab:Creation_of_Test_requirements})
are more significant for the BF and SC algorithms; however, the differences
are not so significant for PG for majority of the test set optimality
criteria. A similar trend can be found for $PL=medium$ although for
the test set optimality criteria $edges_{h}(T)$ and $edges_{m}(T)$,
the differences caused by the atomic and sequence conversions of the
test requirements for the BF and SC algorithms are lower, whereas
this difference is higher for the PG algorithm compared to $PL=high$.

Regarding the \textit{Edge-Pair Coverage} criterion, the situation
changes: the RSC algorithm outperforms the PPT algorithm on all the
test set optimality criteria for $PL=high$ \textcolor{black}{(Table
\ref{tab:Results-Edge-Pair-Coverage}, Figure \ref{fig:Comparison-of-the-algorithms-Edge_PAIR_Coverage-PL_high})
and for $PL=medium$ (Table \ref{tab:Results-Edge-Pair-Coverage},
Figure \ref{fig:Comparison-of-the-algorithms-Edge_PAIR_Coverage-PL_medium}). }

In some cases, the results are relatively similar, for instance scores
for $uedges_{m}(T)$, $e_{h}(T)$ and $e_{m}(T)$ when $PL=high$
and $e_{h}(T)$ and $e_{m}(T)$ when $PL=medium$. However, significant
differences can be observed for the rest of the test set optimality
criteria. \textcolor{black}{For instance, when $PL=high$, the difference
in} $\mid T\mid$ is 39\%, the difference in $edges(T)$ is 25\% and
the difference in $nodes(T)$ is 24\%. When\textcolor{black}{{} $PL=medium$,
the difference in} $\mid T\mid$ is 39\%, the difference in $edges(T)$
is 27\% and the difference in $nodes(T)$ is 25\%.\textcolor{black}{{}
}These results show that the RSC algorithm is a better candidate for
\textit{Edge-Pair Coverage} criterion than is PPT.

Regarding $uedges_{h}(T)$ when $PL=high$ for both Edge Coverage
and Edge-Pair Coverage, all the employed algorithms created the test
sets that had the same value for the $uedges_{h}(T)$ criterion. This
is a correct result and occurs because of the principle behind the
algorithms. The same analogy applies for $uedges_{h}(T)$ and $uedges_{m}(T)$
when $PL=medium$.

Regarding the \textbf{test set selection strategy} (the second part)
other facts can be observed from the data. The most important finding
is that for various problem instances $\mathfrak{\mathcal{G}}$ and
different optimality criteria, different algorithms provide the optimal
test set. This can be observed similarly for \textit{Edge Coverage}
with $PL=high$ (see Figure \ref{fig:Overal-statistics-of-the-TS-selection-mechanism--edge-coverage--PTL-high}
and Table \ref{tab:Results-of-the-TS-selection-mechanism--edge-coverage--PTL-high})
and for \textit{Edge Coverage} with $PL=medium$ (see Figure \ref{fig:Overal-statistics-of-the-TS-selection-mechanism--edge-coverage--PTL-medium}),
\textit{Edge-Pair Coverage} with $PL=high$ (see Figure \ref{fig:Overal-statistics-of-the-TS-selection-mechanism--edge-PAIR-coverage--PTL-high})
and \textit{Edge-Pair Coverage} with $PL=medium$ (see Figure \ref{fig:Overal-statistics-of-the-TS-selection-mechanism--edge-PAIR-coverage--PTL-medium}).
This effect is well documented by a sample of the detailed data provided
in Table \ref{tab:Results-of-the-TS-selection-mechanism--edge-coverage--PTL-high}. 

For certain test set optimality criteria, the algorithms that provide
the optimal solution for all or for the majority of the problem instances
can be identified. For instance, this is true in the case of the \textit{Edge
Coverage} criterion ($PL=high$ and $PL=medium$), using the PPT algorithm
and the test set optimality criteria $\mid T\mid$ and $edges(T)$.
However, for different optimality criteria (e.g., $e_{h}(T)$ and
$e_{m}(T)$), a single algorithm that clearly outperforms the other
algorithms cannot be identified. For $PL=medium$, this effect is
even more obvious and relates to the fact that when $PL=medium$ the
algorithms reflect more priority edges. Moreover, when $PL=medium$,
the data shows that no single algorithm clearly outperforms the others
for the other test set optimality criteria, namely, $uedges(T)$,
$unodes(T)$ and $er(T)$. For $uedges(T)$, for instance, PPT provided
the optimal test set for 31 problem instances, RSC for 16 instances,
BF with atomic conversion of test requirements for 11 instances, BF
with sequence conversion of test requirements for 22 instances, SC
with atomic conversion of test requirements for 14 instances, SC with
sequence conversion of test requirements for 30 instances, PG with
atomic conversion of test requirements for 27 instances and PG with
sequence conversion of test requirements for 33 instances. On a given
problem instance, applying more algorithms can provide an optimal
result.

Generally, the results of the \textit{Edge Coverage} criteria (Figures
\ref{fig:Overal-statistics-of-the-TS-selection-mechanism--edge-coverage--PTL-high}
and \ref{fig:Overal-statistics-of-the-TS-selection-mechanism--edge-coverage--PTL-medium})
correlate with the findings presented when the algorithms were compared
by their average values of optimality criteria (\textcolor{black}{Figures
\ref{fig:Comparison-of-the-algorithms-EdgeCoverage-PL_high}} and
\textcolor{black}{\ref{fig:Comparison-of-the-algorithms-EdgeCoverage-PL_medium}}).

For \textit{Edge-Pair Coverage}, the analysis is simpler, because
only two algorithms, PPT and RSC are comparable for this test coverage
level. When $PL=high$,the RSC outperforms PPT on most of the test
set optimality criteria; however, no clear \char`\"{}winner\char`\"{}
can be identified for criteria $e_{h}(T)$ and $e_{m}(T)$. When considering
$e_{h}(T)$, PPT provided the optimal test set for 25 problem instances,
RSC provided the optimal test set for 33 problem instances, and both
algorithms provided the optimal test set for 8 problem instances.
Considering $e_{m}(T)$ , PPT provided the optimal test set for 27
problem instances, RSC provided the optimal test set for 33 problem
instances, and both algorithms provided the optimal test set for 10
problem instances.

For the optimality criteria $uedges(T)$, $nodes(T)$, $unodes(T)$
and $er(T)$, when $PL=high$, RSC outperformed PPT in 39 out of 50
problem instances , while both algorithms provided the same result
for 6 problem instances.

For $PL=medium$, the RSC outperformed PPT in most of the test set
optimality criteria. For this priority level, RSC yields the better
results. This also applies to the $e_{h}(T)$ and $e_{m}(T)$ previously
discussed for $PL=high$. Considering $e_{h}(T)$ when $PL=medium$,
PPT provided the optimal test set for 18 problem instances, RSC provided
the optimal test set for 36 problem instances, and both algorithms
provided the optimal test set for 4 problem instances. Considering
$e_{m}(T)$ , PPT provided the optimal test set for 19 problem instances,
RSC provided the optimal test set for 36 problem instances, and both
algorithms provided the optimal test set for 4 problem instances.

Generally, the results justify the concept proposed in this paper:
in situations in which different algorithms provide optimal results
for different problem instances (when considering a particular test
set optimality criterion), employing more algorithms and then selecting
the best set is a practical approach.

\section{\label{sec:Threats-to-Validity}Threats to Validity}

Several issues can be raised regarding the validity of the results;
we discuss them in this section and describe the countermeasures that
mitigate the effects of these issues.

The first concern that can be raised involves the generation of the
set of test requirements $R$ from $\mathfrak{\mathcal{G}}$ for the
BF, SC and PG algorithms for the \textit{Edge Coverage} criterion
($TDL=1$), where $PL$ is used for the test set reduction. The SUT
models $\mathfrak{\mathcal{G}}$ and $G$ with $R$ differ between
the methods, how to capture the priority parts of the SUT process,
hence the different possibilities for conversion between the edge
priorities in $\mathfrak{\mathcal{G}}$ and the set of test requirements
$R$ can be discussed. To mitigate this issue, we employed and analyzed
two different strategies for generating the set of test requirements
$R$ from $\mathfrak{\mathcal{G}}$, namely, the atomic and sequence
conversion methods, which are specified in Section \ref{tab:Creation_of_Test_requirements}.

Another issue relates to the topology of the SUT models. The BF, SC
and PG algorithms use a directed graph as the SUT model \cite{li2012better};
consequently, a directed graph is also used for RSC, because RSC employs
SC as its main part (refer to Algorithm \ref{alg:Set-Covering-with-Test-Set-Reduction}).
For PPT, a directed multigraph can be used as input. To mitigate this
issue and to ensure the objective comparability of all the algorithms
and the convertibility of \textbf{$\mathfrak{\mathcal{G}}$} to $G$
and $R$, we used only directed graphs in the experiments.

A related issue arises at this point: Does this restriction not limit
the modeling possibilities when capturing the SUT structure? The answer
is that practically speaking, the modeling possibilities are not limited.
Using a directed graph leads only to more extensive models. When parallel
edges present in the conceptual SUT model (e.g., UML Activity Diagram)
are not allowed in its abstraction as captured by a directed graph,
we instead use graph nodes to capture the parallel edges. This approach
leads to more extensive graphs; however, it does not limit the algorithms
and the overall solution.

Another question can be raised regarding the practical applicability
of all the test set optimality criteria presented in Table \ref{tab:Test-set-optimality-criteria};
many arguments can be brought both for and against this issue. In
this study, rather than tackling such discussions, we present the
data for all the optimality criteria and let the readers decide.

The last issue to be raised regards the strength of the test cases,
which are reduced by the $PL$ concept to cover the priority parts
of the SUT processes only. In these defined priority parts, the test
coverage and the strength of the test cases are guaranteed. However,
it is not guaranteed for the non-priority parts due to the principle
of the $PL$ criteria. However, this fact does not invalidate the
algorithms, the experimental data, or the conclusions drawn from these
data.

\section{\label{sec:Related-Work}Related Work}

In the majority of the current path-based techniques, a SUT abstraction
is based on a directed graph \cite{offutt2008introduction}. To capture
the priority of specific parts of the SUT process or determine the
test coverage level, test requirements are used \cite{offutt2008introduction,shirole2013uml}.
To assess the optimality of a path-based test set, a number of criteria
can be discussed \cite{li2012better,offutt2008introduction,li2009experimental}.
These criteria are usually based on the number of nodes, the number
of edges, the number of paths or the coverage of the test requirements.

To generate path-based test cases, a number of algorithms have been
proposed \cite{dwarakanath2014minimum,li2012better,arora2017synthesizing,sayyari2015automated,hoseini2014automatic,shirole2013uml,anand2013orchestrated},
such as the Brute Force algorithm, the Set-Covering Based Solution,
or the Matching-Based Prefix Graph Solution \cite{li2012better}.
Additionally, genetic algorithms have been employed to generate the
prime paths \cite{hoseini2014automatic} or to generate basis test
paths \cite{ghiduk2014automatic}. Other nature-inspired algorithms
have also been proposed, for example, ant colony optimization algorithms
\cite{sayyari2015automated,srivastava2010optimized}, the firefly
algorithm \cite{srivatsava2013optimal} and algorithms inspired by
microbiology \cite{arora2017synthesizing}.

Test set optimization based on prioritization is considered essential
area to be explored and here; various alternative approaches can be
identified. As an example, clustering based on a neural network was
examined in \cite{gokcce2006coverage}, fuzzy clustering possibilities
were explored in \cite{belli2007coverage}, and the Firefly optimization
algorithm was utilized in \cite{panthi2015generating}. These approaches
also use the internal structure of a SUT as the input to the process.

The path-based testing technique itself is generally applicable to
and can be employed for various types of testing. For instance, the
composition of end-to-end business scenarios \cite{bures2015pctgen},
the composition of scenarios for integration tests or path-based testing
focusing on the code level of the SUT \cite{yan2008efficient,li2009experimental}.
On this last level, path-based testing overlaps with the data-flow
technique, which focuses on verifying the data consistency of the
SUT code \cite{chaim2013efficient,denaro2014right,denaro2015dynamic,su2017survey}.
In this area, control-flow graphs are employed as the SUT abstraction
\cite{yan2008efficient}.

Alternative approaches to the current test requirement concept have
been formulated \cite{bures2017prioritized}, which result in capturing
the priorities by the weights of the graph edges. This approach was
inspired by the need for more priority levels, which are commonly
used in the software engineering and management praxes \cite{achimugu2014systematic,van2013tmap}.
Another motivation for this approach regards certain limitations of
the test requirements concept: in a number of the algorithms, the
test requirements can be practically used either to specify the SUT
priority parts or to determine the test intensity. As an alternative,
the PPT was formulated, which is an algorithm that combines variable
test coverage with SUT part prioritization \cite{bures2017prioritized}. 

Regarding using a combination of algorithms to determine the optimal
test set, significantly less work exists. Some work utilizing this
idea exists in the area of combinatorial interaction testing, in which
different approaches are combined to obtain the optimal test set \cite{zamli2016tabu,zamli2017experimental}.
Considering the experimental results presented in this paper, this
stream can be considered prospectively for the path-based testing
domain.

\section{Conclusion\label{sec:Conclusion}}

In the paper, we proposed a strategy that employs a set of currently
available algorithms and one new algorithm to find an optimal set
of path-based test cases for a SUT model based on a directed graph
with priority parts. The priority is captured as edge weights; for
some of the algorithms, it is converted to test requirements. The
optimality of the test set is determined by an optimality criterion
selected by the user from fourteen indicators of test set optimality,
by an optimality function that can be parameterized, or by the sequence
selection method specified in this paper. The experimental results
from running this strategy on 50 various problem instances justify
the proposed approach. For the various problem instances and different
optimality criteria, different algorithms provide the optimal test
set\textemdash an outcome that was observed for all four combinations
of test coverage and priority level criteria used in the experiments. 

From the exercised algorithms, the PPT provided the best results for
the \textit{\textcolor{black}{Edge Coverage}} criterion. However,
for certain sets of problem instances and certain test set optimality
criteria, the PPT is outperformed by other algorithms (i.e., RSC,
SC, and PG) and by BF in certain instances. For the \textit{Edge-Pair
Coverage} criterion, where the PPT and the RSC were the only comparable
candidates for solving the problem (combining \textit{Edge-Pair Coverage}
with prioritization of particular SUT model parts), the RSC outperformed
the PPT on the majority of the optimality criteria. However, for specific
optimality criteria (e.g., $e_{h}(T)$ and $e_{m}(T)$), the dominance
of the RSC algorithm was weak, and for a significant proportion of
the problem instances, the PPT provided better results.

The proposed test set selection strategy is not a substitute for the
development of new perspective algorithms to solve the path-based
test case generation problem. Using this strategy, the quality of
the overall result depends on the quality of the algorithms employed.
If new algorithms are developed that provide better results for particular
problem instances, this strategy could provide better results in the
future.

\section*{Acknowledgments}

This research is conducted as a part of the project TACR TH02010296
Quality Assurance System for the Internet of Things Technology.

\bibliographystyle{IEEEtran}
\bibliography{13C___cvut_habilitace_arxiv_org_PCT_paper_mybibfile}

\end{document}